\documentclass[sigconf]{acmart}

\usepackage{amsmath}
\usepackage{subfigure}
\usepackage{multirow}

\AtBeginDocument{%
  \providecommand\BibTeX{{%
    \normalfont B\kern-0.5em{\scshape i\kern-0.25em b}\kern-0.8em\TeX}}}

\copyrightyear{2022} 
\acmYear{2022} 
\setcopyright{acmcopyright}\acmConference[WSDM '22]{Proceedings of the Fifteenth ACM International Conference on Web Search and Data Mining}{February 21--25, 2022}{Tempe, AZ, USA}
\acmBooktitle{Proceedings of the Fifteenth ACM International Conference on Web Search and Data Mining (WSDM '22), February 21--25, 2022, Tempe, AZ, USA}
\acmPrice{15.00}
\acmDOI{10.1145/3488560.3498433}
\acmISBN{978-1-4503-9132-0/22/02}

\begin{document}
\fancyhead{}
\title{Contrastive Learning for Representation Degeneration Problem in Sequential Recommendation}

\author{Ruihong Qiu, Zi Huang, Hongzhi Yin*, and Zijian Wang}
\affiliation{%
  \institution{The University of Queensland}
  \city{Brisbane}
  \country{Australia}
}
\email{{r.qiu, h.yin1, zijian.wang}@uq.edu.au, huang@itee.uq.edu.au}

\begin{abstract}
Recent advancements of sequential deep learning models such as Transformer and BERT have significantly facilitated the sequential recommendation. However, according to our study, the distribution of item embeddings generated by these models tends to degenerate into an anisotropic shape, which may result in high semantic similarities among embeddings. In this paper, both empirical and theoretical investigations of this representation degeneration problem are first provided, based on which a novel recommender model DuoRec is proposed to improve the item embeddings distribution. Specifically, in light of the uniformity property of contrastive learning, a contrastive regularization is designed for DuoRec to reshape the distribution of sequence representations. Given the convention that the recommendation task is performed by measuring the similarity between sequence representations and item embeddings in the same space via dot product, the regularization can be implicitly applied to the item embedding distribution. Existing contrastive learning methods mainly rely on data level augmentation for user-item interaction sequences through item cropping, masking, or reordering and can hardly provide semantically consistent augmentation samples. In DuoRec, a model-level augmentation is proposed based on Dropout to enable better semantic preserving. Furthermore, a novel sampling strategy is developed, where sequences having the same target item are chosen hard positive samples. Extensive experiments conducted on five datasets demonstrate the superior performance of the proposed DuoRec model compared with baseline methods. Visualization results of the learned representations validate that DuoRec can largely alleviate the representation degeneration problem.
\end{abstract}

\begin{CCSXML}
<ccs2012>
<concept>
<concept_id>10002951.10003317.10003347.10003350</concept_id>
<concept_desc>Information systems~Recommender systems</concept_desc>
<concept_significance>500</concept_significance>
</concept>
</ccs2012>
\end{CCSXML}

\ccsdesc[500]{Information systems~Recommender systems}

\keywords{sequential recommendation, contrastive learning}

\maketitle
\let\thefootnote\relax\footnote{* Corresponding author.}

\begin{figure}[h]
    \centering
    \subfigure[SASRec.]{
    \label{fig:svd-sasrec-clothing}
    \includegraphics[width=0.261\linewidth]{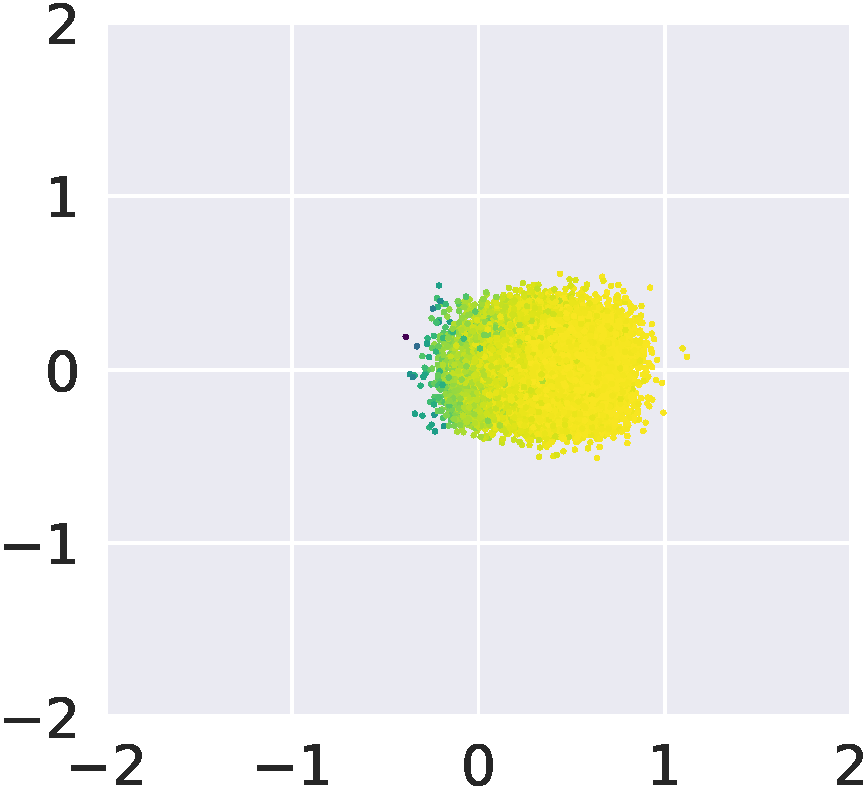}
    }
    \subfigure[DuoRec.]{
    \label{fig:svd-USCLRec-clothing}
    \includegraphics[width=0.322\linewidth]{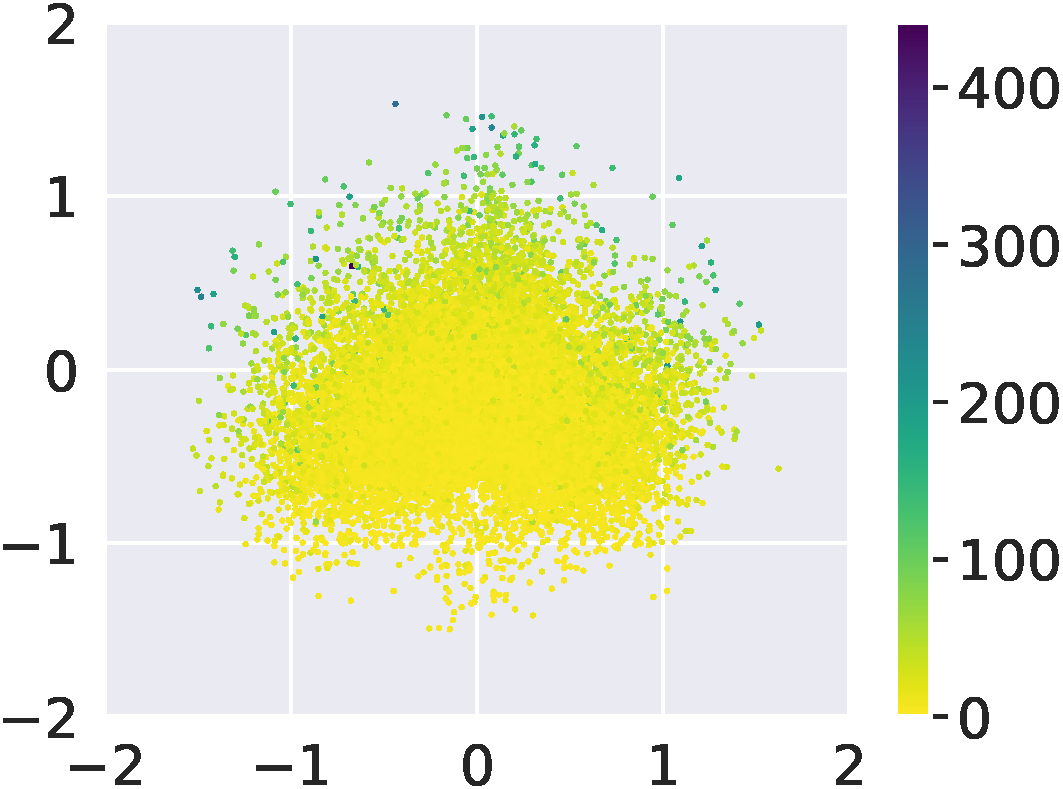}
    }
    \subfigure[Singular values.]{
    \label{fig:svs}
    \includegraphics[width=0.33\linewidth]{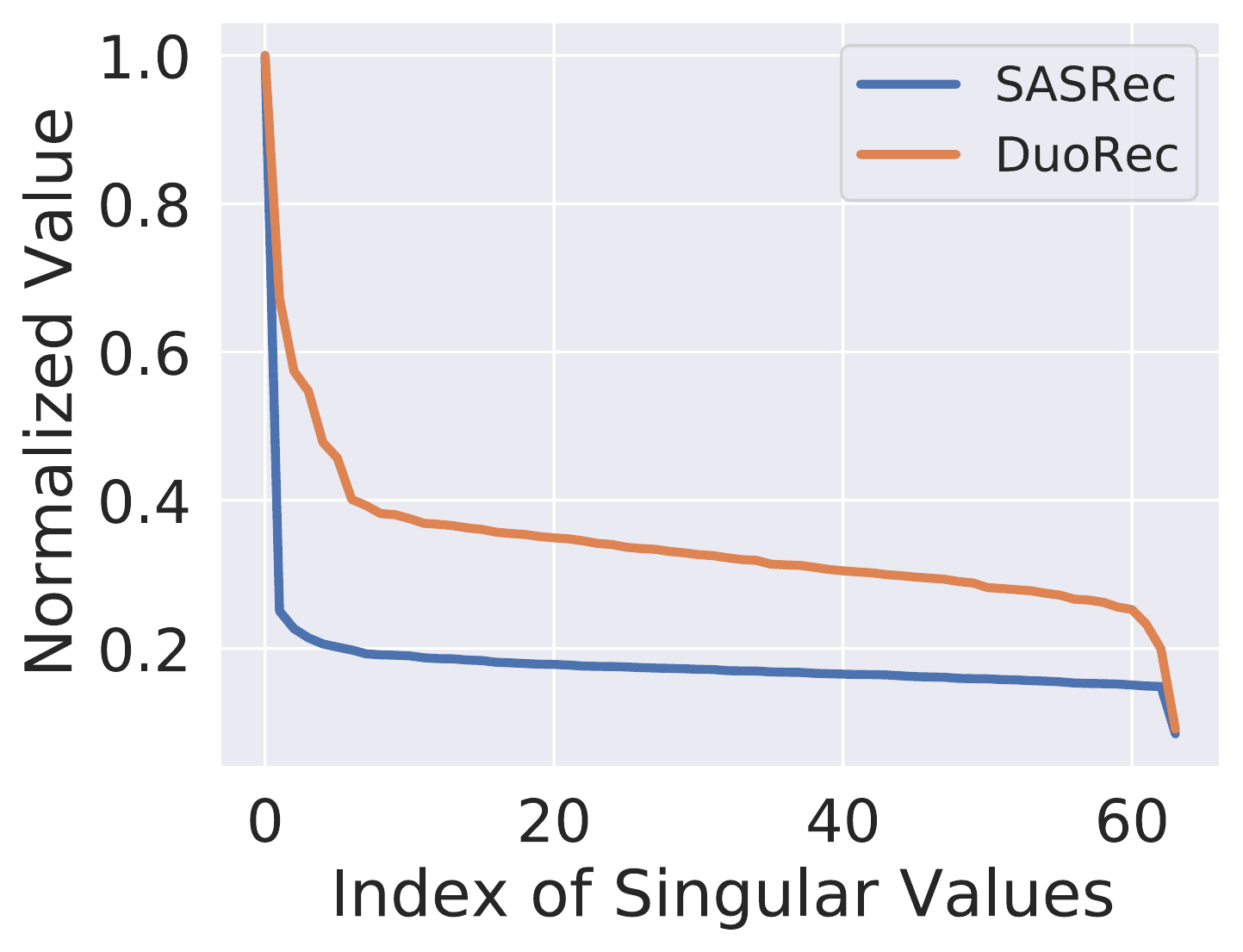}
    }
    \caption{The item embedding matrix of the Amazon Clothing dataset is projected into 2D by SVD with colors indicating the frequency of items in the dataset. (a) Item embeddings learned by SASRec. Most of the rare items fall into a narrow cone, leading to high similarities among one another due to geometric properties. (b) Item embeddings learned by DuoRec. The distribution of item embeddings is more uniform in terms of both the magnitude and the frequency. (c) Normalized singular values of item embedding matrices. The fast decrease singular values of SASRec indicate the item embedding matrix is approximately in extreme low-rank. The slow decrease singular values of DuoRec reflect that the item embeddings are more representative.}
    \label{fig:svd-problem}
\end{figure}

\section{Introduction}
Traditional recommender systems usually predict a user's preference based on their historical records without considering the time factor~\cite{bprmf,fpmc,item-knn,causalrec} while the preference generally shifts as time goes on. Recent sequential recommendation methods exploit sequential patterns of the user's interactions to capture the dynamic preference~\cite{gru4rec,caser,sasrec,bert4rec,safm,s3rec,s2s,cl4srec,mminforec,lightweight,seq2graph,fgnn,fgnnj,gag,posrec}.

During our study of these sequential models, a representation degeneration problem is observed in the item embeddings, whose distribution degenerates into a narrow cone and leads to an indiscriminate representation of the semantics. As shown in Figure~\ref{fig:svd-sasrec-clothing}, the item embeddings generated by SASRec~\cite{sasrec} tend to be positive along the X-axis while distributing narrowly in the Y-axis direction. In this situation, most of the items are positively related to one another due to the geometric properties. This distribution is typically anisotropic~\cite{Contextual,bert-flow,cosreg,sc,too-much}, reflected by which the singular values of the item embedding matrix quickly decrease to small values in the blue curve of Figure~\ref{fig:svs}. There is a dominant dimension in the embedding matrix while other dimensions are ineffective, which is approximately in extreme low rank. In contrast, item embeddings of the proposed DuoRec are distributed more uniformly around the origin point with the singular values decreasing more slowly as in Figure~\ref{fig:svd-USCLRec-clothing} and the orange curve in Figure~\ref{fig:svs}.

In this paper, we first investigate the causes of the representation degeneration with theoretical analysis, motivated by which a novel sequential recommender model DuoRec is proposed to improve the distribution of item embeddings. Specifically, inspired by the uniformity property of contrastive learning, a contrastive regularization is first designed to enhance the uniformity of the sequence representation distribution. Given that the recommendation is generally performed by measuring the similarity between sequence representations and item embeddings in the same space via dot product, the contrastive regularization can implicitly influence the item embeddings to distribute more uniformly. Existing contrastive learning methods generally generate positive samples using data-level augmentation, e.g., item cropping, masking, and reordering~\cite{cl4srec,consert}, which may cause semantically inconsistent samples. Considering that the input data itself is generally embedded into a dense vector, we propose a model-level augmentation method, which applies two different sets of Dropout masks to the sequence representation learning. Moreover, since there are a large number of semantically similar sequences representing similar user preferences, an extra positive sampling strategy is developed to generate hard and informative positive samples, where sequences with the same target item are considered semantically similar.

The main contributions of this paper are summarized as follows:
\begin{itemize}
    \item The representation degeneration problem is identified and investigated in sequential recommender models with theoretical and empirical analysis.
    \item To address the representation degeneration problem, a novel model DuoRec is proposed with a contrastive objective serving as the regularization over sequence representations.
    \item A model-level augmentation for user sequences is designed based on Dropout. Furthermore, a positive sampling strategy is developed using the target item as the supervision signal.
    \item Extensive experiments are conducted on five benchmark datasets, which show the state-of-the-art performance of the proposed DuoRec model and the effectiveness of the contrastive regularization for sequential recommendation.
\end{itemize}

\section{Representation Degeneration Problem}
\label{sec:seq-emb}
\subsection{Notations and Task Definition}
\label{sec:notation}
In sequential recommendation, the problem setting is using historical interactions to infer user's preference and recommend the next item. There is an item set $\mathcal{V}$ containing all items and $| \mathcal{V}|$ is the number of items. The historical interactions of a user are constructed as an ordered list $s=\left[v_1,v_2,\ldots,v_t \right]$, where $v_i\in\mathcal{V},0\leq i\leq t$ and $t$ indicates the current time step as well as the length of $s$. The recommendation task is to predict the next item at time step $t+1$, i.e., $v_{t+1}$ for the user. In the following, bold lowercase and uppercase symbols are used to denote vectors and matrices respectively.

\subsection{Representation Degeneration Problem}
\label{sec:seq-learning}
To perform the next item prediction task in sequential recommendation, an interaction sequence is encoded into a fixed-length vector by the model to conduct a retrieval over the item set.

Given a sequence of items $s=\left[v_1,v_2,\ldots,v_t \right]$, the model calculates the probability of this sequence as $p(s)=\prod^t_{n=1}p(v_n|c_n)$, where $c_n$ is the context of an interaction at time step $n$, containing all the previous interactions $v_{<n}$. The $\log$ probability of the sequence can be represented with a $\theta$ parameterized model:
\begin{equation}
\log p_{\theta}(s)=\sum_{n=1}^{t} \log p_{\theta}\left(v_n \mid c_n\right)=\sum_{n=1}^{t} \log \frac{\exp \left(\left\langle\boldsymbol{h}_{c}, \boldsymbol{v_n}\right\rangle\right)}{\sum_{\boldsymbol{v'}}^{\mathcal{V}} \exp \left(\left\langle\boldsymbol{h}_{c}, \boldsymbol{v'}\right\rangle\right)},
\end{equation}
where $\boldsymbol{h_c}\in\mathbb{R}^d$ is the $d$-dimension vector of the context and $\boldsymbol{v_n},\boldsymbol{v'}\in\mathbb{R}^d$ are the embeddings of item. Generally, the context is generated by sequential models such as GRU~\cite{gru} and Transformer~\cite{attention}.

When using the cross-entropy loss to optimize the parameterized model above, the objective function can be abstracted as:
\begin{equation}
\label{eq:ar-obj}
    J=-\mathbb{E}_{s \sim p_{\operatorname{data}}}\left[\log p_{\theta}(s)\right].
\end{equation}

According to~\cite{pmi,bert-flow}, in a well trained sequential model, the dot product term can be approximately decomposed as:
\begin{equation}
\label{eq:pmi}
    \left\langle\boldsymbol{h}_{c}, \boldsymbol{v_n}\right\rangle\approx\log p(v_n|c_n)+\lambda_{c_n}=\text{PMI}(v_n,c_n)+\log p(v_n)+\lambda_{c_n},
\end{equation}
where $p(\cdot)$ is the true probability, $\text{PMI}(v_n,c_n)=\log\frac{p(v_n,c_n)}{p(v_n)p(c_n)}$ is the pointwise mutual information (PMI) between $v_n$ and $c_n$, and $\lambda_{c_n}$ is a context-related term. PMI captures the co-occurrence statistics between the variables, which is usually considered as semantics between tokens and the context.

To optimize the model with Equation (\ref{eq:ar-obj}), the gradient of the loss function with respect to an item embedding $\boldsymbol{v^*}$ that does not appear in the input sequence is:
\begin{equation}
\label{eq:gradient}
    \frac{\partial J}{\partial \boldsymbol{v^*}}=-\frac{\exp \left(\left\langle\boldsymbol{h}_{c}, \boldsymbol{v^*}\right\rangle\right)}{\sum_{\boldsymbol{v'}}^{\mathcal{V}} \exp \left(\left\langle\boldsymbol{h}_{c}, \boldsymbol{v'}\right\rangle\right)}\boldsymbol{h_c}\approx-p(v^*|c_n)\boldsymbol{h_c}.
\end{equation}
This gradient means that for those items with lower frequency in the dataset, the gradient direction is almost determined by the context vector. This is also reflected by Figure~\ref{fig:svd-sasrec-clothing}, where the yellow dots, representing items with a lower frequency, move in a similar direction within a narrow space. This is because most of the time, these item embeddings are trained with the gradient in Equation (\ref{eq:gradient}) as non-target items rather than being target items and trained with the gradient flowing through the encoder. Such a distribution of embeddings is referred to as an anisotropic space~\cite{bert-flow,cosreg}.

As Equation (\ref{eq:pmi}) indicates, the semantic information between the sequence embedding and the item embedding is captured by dot product based on the co-occurrence. However, according to Equation (\ref{eq:gradient}), it is impractical to expect the sequence embeddings can present a clear difference in measuring the similarity between the rare items. Generally, the output layer of a recommender system is a dot product between the sequence representation and the item embeddings, which sets sequences and items representations in the same latent space. In the following sections, a contrastive regularization will be introduced to relocate the distribution of sequence representations around the origin point, which implicitly improves the distribution of item embeddings.

\section{Preliminary: Contrastive Learning}
\label{sec:prelim}
\subsection{Noise Contrastive Estimation}
\label{sec:nce}
Contrastive learning is a training scheme that pulls the positive pairs of samples closer and pushes the negative pairs of samples away~\cite{nce,nce2}. Specifically, a Noise Contrastive Estimation (NCE) objective is generally applied to train an encoder $f$:
\begin{equation}
\label{eq:nce}
    \ell_{\text{NCE}}=\underset{\substack{(x, x^+) \sim p_{\text {pos }}\\x_{i}^{-}\stackrel{\text{i.i.d.}}{\sim} p_{\text {data }}}}{\mathbb{E}}\left[-\log \frac{e^{f(x^+)^{\top} f(x) / \tau}}{e^{f(x^+)^{\top} f(x) / \tau}+\sum_{i} e^{f(x_{i}^{-})^{\top} f(x) / \tau}}\right],
\end{equation}
where $x$ and $x^+$ are a pair of semantically close samples from the distribution $p_\text{pos}$, which serve as the positive pair here while $x$ and $x_i^-$ are a pair of negative samples and $x_i^-$ is sampled randomly from $p_\text{data}$. $\tau$ is the temperature parameter.

\subsection{Alignment and Uniformity}
\label{sec:align}
The NCE loss is intuitively performing the push and pull game according to Equation (\ref{eq:nce}). The mathematical descriptions are formally defined as the alignment and the uniformity of the representations under the assumption that vectors are normalized~\cite{align}:
\begin{align}
    \ell_{\text {align}} &\triangleq \underset{\left(x, x^+\right) \sim p_{\text {pos}}}{\mathbb{E}}\left\|f(x^+)-f\left(x\right)\right\|^{2},\label{eq:align}\\
    \ell_{\text {uniform}} &\triangleq \log \underset{ x^-\stackrel{\text{i.i.d.}}{\sim}p_{\text {data}}}{\mathbb{E}} e^{-2\|f(x^-)-f(x)\|^{2}},\label{eq:uniform}
\end{align}
where $p_\text{pos}$ denotes the distribution of the positive pair of samples and $p_\text{data}$ is the distribution of the independent samples.

For Equation (\ref{eq:align}), minimizing $\ell_{\text {align }}$ is equivalent to encourage the learned representations of $x$ and $x^+$ from a positive pair distribution $p_{\text{pos}}$ to be close. For Equation (\ref{eq:uniform}), minimizing $\ell_{\text {uniform }}$ is equivalent to encourage the uniform distribution of the representations of all the data samples from the distribution $p_{\text{data}}$.

\section{Method: DuoRec}
\label{sec:method}
\subsection{Sequence Encoding As User Representation}
\label{sec:seq-enc}
In sequential recommendation, the main idea is to aggregate the historical interactions to profile the user's preference. Similar to SASRec, the encoding module of DuoRec is a Transformer~\cite{attention}. To leverage the encoding ability of Transformer, the items are firstly converted into embeddings. Then a multi-head self-attention module is applied to compute the user representation.

\subsubsection{Embedding Layer}
\label{sec:el}
In DuoRec, there is an embedding matrix $\boldsymbol{V} \in \mathbb{R}^{|\mathcal{V}| \times d}$, where $d$ is the dimension of the embedding. For the input sequence $s=[v_1,v_2,\ldots,v_{t}]$, the embedding representations are $\boldsymbol{s}=[\boldsymbol{v}_1, \boldsymbol{v}_2,\ldots,\boldsymbol{v}_t]$, where $\boldsymbol{v}_*$ is the embedded vector.

To preserve the time order of the sequence, a positional encoding matrix $\boldsymbol{P}\in \mathbb{R}^{N \times d}$ is constructed, where $N$ indicates the maximum length of all the sequences. Formally, the item embedding and the positional encoding are added up as the input vector for the interaction at a time step $t$ of the Transformer:
\begin{equation}
\label{eq:emb}
\boldsymbol{h}^0_t=\boldsymbol{v}_t+\boldsymbol{p}_t,
\end{equation}
where $\boldsymbol{h}^0_t\in\mathbb{R}^d$ is the complete input vector of the interaction at $t$ and $\boldsymbol{p}_t$ is the positional encoding of the time step $t$.

\subsubsection{Self-attention Module}
\label{sec:mh}
After obtaining the input sequences, the Transformer is applied to compute the updated representations of each item by the multi-head attention mechanism~\cite{attention}. Assuming $\boldsymbol{H}^0=\left[\boldsymbol{h}^0_0,\boldsymbol{h}^0_1,\ldots,\boldsymbol{h}^0_t\right]$ is the hidden representation of the sequence as both the input of an $L$-layer multi-head Transformer encoder (Trm), the encoding procedure of the sequence can be defined as:
\begin{equation}
    \boldsymbol{H}^L=\operatorname{Trm}\left(\boldsymbol{H}^0\right),
\end{equation}
where the last hidden vector $\boldsymbol{h}^L_t$ in $\boldsymbol{H}^L=\left[\boldsymbol{h}^L_0,\boldsymbol{h}^L_1,\ldots,\boldsymbol{h}^L_t\right]$ is selected to be the user representation of this user sequence.

\subsection{Recommendation Learning}
\label{sec:learn}
The next item prediction task is framed as a classification task over the whole item set. Given the sequence representation $\boldsymbol{h}$ and the item embedding matrix $\boldsymbol{V}$, the predictive score is calculated as:
\begin{equation}
\label{eq:dp}
    \hat{\boldsymbol{y}}=\operatorname{softmax}\left(\boldsymbol{V}\boldsymbol{h}\right),
\end{equation}
where $\hat{\boldsymbol{y}}\in\mathbb{R}^{|\mathcal{V}|}$. With the index of the ground truth item converted into a one-hot vector $\boldsymbol{y}$, the cross-entropy loss is calculated as:
\begin{equation}
    \ell_\text{Rec}=-\text{one-hot}(\boldsymbol{y}_i)\log(\hat{\boldsymbol{y}_i}).
\end{equation}

\subsection{Contrastive Regularization}
\label{sec:regularization}
To alleviate the representation degeneration problem, a contrastive regularization is developed by the exploitation of both the unsupervised and the supervised contrastive samples.

\subsubsection{Unsupervised Augmentation}
The unsupervised contrastive augmentation in DuoRec aims to provide a semantically meaningful augmentation for individual sequences in an unsupervised style. In the previous method such as CL4SRec~\cite{cl4srec}, the augmentation methods include item cropping, masking, and reordering. Similar techniques in natural language processing are applied, e.g., word deletion, reordering, and substitution~\cite{clear,cocolm}. Although these methods provide augmentations that help to improve the performance of the corresponding models to some extent, the augmentations cannot provide a guarantee for high semantic similarity. Since the data-level augmentation is not perfectly fit for discrete sequence, a model-level augmentation is proposed in this paper. In the computation of a sequence vector, there are Dropout modules in both the embedding layer and the Transformer encoder. Forward-passing an input sequence twice with different Dropout masks will generate two different vectors, which are semantically similar while having different features. Therefore, we choose a different Dropout mask for the unsupervised augmentation of the input sequence $s$, which is firstly operated on the input embedding to the Transformer encoder in Equation (\ref{eq:emb}) to obtain an $\boldsymbol{h}^{0'}_t$. Afterward, the augmented input sequence embedding is fed into the Transformer encoder with the same weight but a different Dropout mask:
\begin{equation}
    \boldsymbol{H}^{L'}=\operatorname{Trm}(\boldsymbol{H}^{0'}),\quad\boldsymbol{h}'=\boldsymbol{h}_t^{L'}=\boldsymbol{H}^{L'}[-1],
\end{equation}
where $[-1]$ is mimicking the Python style of indexing the last element in the list and $\boldsymbol{h}'$ is the augmented sequence representation.

\subsubsection{Supervised Positive Sampling}
The supervised contrastive augmentation in DuoRec aims to incorporate the semantic information between semantically similar sequences into the contrastive regularization. The reason why semantic positives are required is that if only unsupervised contrastive learning is applied, the originally semantic similar samples will be categorized into the negative samples~\cite{scl}. Therefore, the most important thing is to determine what samples are semantically similar.

\paragraph{Semantic Similarity}
In sequential recommendation, the goal is to predict users' preferences. If two sequences represent the same user preference, it is natural to infer these two sequences that contain the same semantics. Therefore, given two different user sequences $s_i=\left[v_{i,1},v_{i,2},\ldots,v_{i,t^i}\right]$ and $s_j=\left[v_{j,1},v_{j,2},\ldots,v_{j,t^j}\right]$, if the predictive objectives of $s_i$ and $s_j$, i.e., $v_{i,t^i+1}$ and $v_{j,t^j+1}$, are the same item, $s_i$ and $s_j$ are considered semantically similar in DuoRec.

\paragraph{Positive Sample}
For the input sequence $s$, there are sequences having the same target item in the dataset. A semantic similar sequence $s_s$ is randomly sampled from these sequences. With the input embedding $\boldsymbol{H}^{0'}_s$, the supervised augmentation is:
\begin{equation}
    \boldsymbol{H}^{L'}_s=\operatorname{Trm}(\boldsymbol{H}^{0'}_s),\quad\boldsymbol{h}'_s=\boldsymbol{h}_{t,s}^{L'}=\boldsymbol{H}^{L'}_s[-1],
\end{equation}
where $\boldsymbol{h}'_s$ is the augmented sequence representation.

\subsubsection{Negative Sampling}
To effectively construct the negative samples for an augmented pair of samples, all the other augmented samples in the same training batch are considered negative samples. Assuming that the training batch is $\mathcal{B}$ and the batch size is $|\mathcal{B}|$, after the augmentation, there will be $2|\mathcal{B}|$ hidden vectors, $\left\{\boldsymbol{h}'_1,\boldsymbol{h}'_{1,s},\boldsymbol{h}'_2,\boldsymbol{h}'_{2,s},\ldots,\boldsymbol{h}'_{|\mathcal{B}|},\boldsymbol{h}'_{|\mathcal{B}|,s}\right\}$, where the subscript and superscript are overloaded to denote the index of samples in the batch and the augmentations for clarity. Therefore, for each positive pair in the batch, there are $2(|\mathcal{B}|-1)$ negative pairs as the negative set $\mathcal{S}^-$. For example, for the augmented pair of sequence representations $\boldsymbol{h}'_1$ and $\boldsymbol{h}'_{1,s}$, the corresponding negative set $\mathcal{S}_1^-=\left\{\boldsymbol{h}'_2,\boldsymbol{h}'_{2,s},\boldsymbol{h}'_3,\boldsymbol{h}'_{3,s},\ldots,\boldsymbol{h}'_{|\mathcal{B}|},\boldsymbol{h}'_{|\mathcal{B}|,s}\right\}$. If there are sequences with the same target item, these sequences will be removed from $\mathcal{S}^-$ as well.

\subsubsection{Regularization Objective}
Similar to Equation (\ref{eq:nce}), the contrastive regularization for the batch $\mathcal{B}$ in DuoRec is defined as:
\begin{align}
\label{eq:reg}
    \ell_{\text{Reg}}&=\underset{i\in|\mathcal{B}|}{\mathbb{E}}\left[-\log \frac{e^{(\boldsymbol{h}'_{i})^{\top} (\boldsymbol{h}'_{i,s}) / \tau}}{e^{(\boldsymbol{h}'_{i})^{\top} (\boldsymbol{h}'_{i,s}) / \tau}+\sum\limits_{\boldsymbol{s^-}\in\mathcal{S}_i^-} e^{(\boldsymbol{h}'_{i})^{\top} (\boldsymbol{s^-}) / \tau}}\right]\notag\\
    &+\underset{i\in|\mathcal{B}|}{\mathbb{E}}\left[-\log \frac{e^{(\boldsymbol{h}'_{i,s})^{\top} (\boldsymbol{h}'_{i}) / \tau}}{e^{(\boldsymbol{h}'_{i,s})^{\top} (\boldsymbol{h}'_{i}) / \tau}+\sum\limits_{\boldsymbol{s^-}\in\mathcal{S}_i^-} e^{(\boldsymbol{h}'_{i,s})^{\top} (\boldsymbol{s^-}) / \tau}}\right],
\end{align}
which computes twice for the unsupervised and the supervised augmented representation respectively.

Thus, the overall objective of DuoRec with $\lambda$ scale weight is:
\begin{equation}
\label{eq:all-loss}
    \ell=\ell_\text{Rec}+\lambda\ell_\text{Reg}.
\end{equation}

\subsection{Discussion}
\label{sec:discussion}
In this section, the properties of the contrastive regularization of DuoRec and the connection with other methods will be described.

\subsubsection{Solving Representation Degeneration Problem}
To investigate how the contrastive regularization can solve the representation degeneration problem, the property of the contrastive regularization $\ell_{\text{Reg}}$ in Equation (\ref{eq:reg}) needs to be analyzed. According to Equation (\ref{eq:align}) and (\ref{eq:uniform}), the alignment and the uniformity of $\ell_{\text{Reg}}$ are as follows:
\begin{align}
        \ell_{\text {Reg,align}} &\triangleq \underset{\left(\boldsymbol{h}'_i,\boldsymbol{h}'_{i,s}\right) \sim p_{\text {pos}}}{\mathbb{E}}\left\|\boldsymbol{h}'_{i}-\boldsymbol{h}'_{i,s}\right\|^{2},\label{eq:reg-align}\\
    \ell_{\text {Reg,uniform,first}} &\triangleq \log \underset{\boldsymbol{s} \stackrel{\text{i.i.d.}}{\sim} p_{\text {data}}}{\mathbb{E}} e^{-2\|\boldsymbol{h}'_{i}-\boldsymbol{s}\|^{2}},\label{eq:reg-uniform-1}\\
    \ell_{\text {Reg,uniform,second}} &\triangleq \log \underset{\boldsymbol{s} \stackrel{\text{i.i.d.}}{\sim} p_{\text {data}}}{\mathbb{E}} e^{-2\|\boldsymbol{h}'_{i,s}-\boldsymbol{s}\|^{2}}.\label{eq:reg-uniform-2}
\end{align}

In the alignment term, it is meaningful to keep the alignment between the positive pairs of representations from two augmentations of the same input sequence. While in the uniformity term, the objective is to uniformly distribute the representations of the sequences. The alignment between semantic positive pairs is pulling the representations of semantically similar sequences together. While the uniformity term is pushing all the sequence representations to be uniformly distributed. Since the main learning objective of recommendation is performed by the dot product between the sequence representation and the item embeddings in Equation (\ref{eq:dp}), it is meaningful to regularize the distribution of the sequence representation so that the distribution of item embeddings can be influenced.

In the representation degeneration problem, an essential drawback of the cone distribution is that there is a dominant axis of the embeddings. Based on the uniformity, this situation will be eased because the sequence representation will be distributed uniformly, which will guide the distribution of the item embeddings via the dot product in Equation (\ref{eq:dp}). While for the other drawback that rare words tend to be far away from the origin point, $\ell_{\text{Reg}}$ alleviates this phenomenon by adjusting the gradient for rare words. According to the analysis of the gradient in Equation (\ref{eq:gradient}) for rare words, the uni-direction of the gradient is because most of the rare words are mainly trained with the recommendation softmax loss. With the contrastive regularization, these rare words are exposed more often than before since there will be more positive and negative samplings, which are trained via the gradient flow through the encoder rather than directly on the embeddings.

\subsubsection{Connection}
Recent methods use the contrastive objective mainly for regularization. For example, CL4SRec~\cite{cl4srec} augments the input sequence in data-level with cropping, masking, and reordering. This is directly following the normal contrastive paradigm for computer vision to augment the samples in the input space. However, discrete sequences are hard to determine the semantic content and even harder to provide a semantically consistent augmentation. If the unsupervised Dropout augmentation of DuoRec is operated twice and only these unsupervised augmented samples are used for the contrastive regularization, it becomes the Unsupervised Contrastive Learning (UCL) variant in the following experiment. Since the augmentation of UCL avoids the data-level augmentations, which cannot guarantee the augmented samples still contain similar semantics, the UCL can outperform CL4SRec consistently. Similarly, if only the supervised augmentation of DuoRec is used, then it becomes the Supervised Contrastive Learning (SCL) variant, which can provide a harder training objective. And SCL can outperform UCL for using more appropriate samples. This is also observed by recent natural language processing research~\cite{simcse}.

\section{Experiment}
\label{sec:exp}
In experiments, we will answer these research questions (RQ):
\begin{itemize}
    \item \textbf{RQ1}: How does the DuoRec perform compared with the state-of-the-art methods? (Section~\ref{sec:rq1})
    \item \textbf{RQ2}: How does the DuoRec perform compared with the existing contrastive training paradigms? (Section~\ref{sec:rq-cl})
    \item \textbf{RQ3}: How does contrastive regularization help with the training? (Section~\ref{sec:rq-vis-cl})
    \item \textbf{RQ4}: How is the sensitivity of the hyper-parameters in the DuoRec model? (Section~\ref{sec:rq-param})
\end{itemize}

\begin{table}[t]
    \caption{Statistics of the datasets after preprocessing.}
    \small
    \begin{tabular}{l|rrrrr}
    \toprule
    Specs. & Beauty & Sports & Clothing & ML-1M & Yelp \\
    \midrule
    $\sharp$ Users & 22,363 & 35,598 & 39,387 & 6,041 & 30499\\
    $\sharp$ Items & 12,101 & 18,357 & 23,033 & 3,417 & 20068\\
    $\sharp$ Avg. Length & 8.9 & 8.3 & 7.1 & 165.5 & 10.4\\
    $\sharp$ Actions & 198,502 & 296,337 & 278,677 & 999,611 & 317182\\
    Sparsity & 99.93\% & 99.95\% & 99.97\% & 95.16\% & 99.95\%\\
    \bottomrule
    \end{tabular}
    \label{tab:datasets}
\end{table}

\begin{table*}[t]
    \caption{Overall performance. Bold scores represent the highest results of all methods. Underlined scores stand for the highest results from previous methods. The DuoRec achieves the state-of-the-art result among all baseline models.}
    \small
    \begin{tabular}{c|l|ccccccc|c|c}
    \toprule
         Dataset& Metric & BPR-MF & GRU4Rec & Caser & SASRec & BERT4Rec &$\text{S}^3\text{Rec}_\text{MIP}$ &CL4SRec& DuoRec & Improv.\\
         \midrule
         \multirow{4}*{Beauty}&HR@5&0.0120&0.0164&0.0259&0.0365&0.0193&0.0327&\underline{0.0401}&\textbf{0.0546}$\pm$0.0013&35.91\%\\
         &HR@10&0.0299&0.0365&0.0418&0.0627&0.0401&0.0591&\underline{0.0683}&\textbf{0.0845}$\pm$0.0010&19.17\%\\
         &NDCG@5&0.0040&0.0086&0.0127&0.0236&0.0187&0.0175&\underline{0.0223}&\textbf{0.0352}$\pm$0.0006&57.85\%\\
         &NDCG@10&0.0053&0.0142&0.0253&0.0281&0.0254&0.0268&\underline{0.0317}&\textbf{0.0443}$\pm$0.0006&39.75\%\\
         \midrule
         \multirow{4}*{Clothing}&HR@5&0.0067&0.0095&0.0108&\underline{0.0168}&0.0125&0.0163&\underline{0.0168}&\textbf{0.0193}$\pm$0.0012&14.88\%\\
         &HR@10&0.0094&0.0165&0.0174&\underline{0.0272}&0.0208&0.0237&0.0266&\textbf{0.0302}$\pm$0.0009&11.03\%\\
         &NDCG@5&0.0052&0.0061&0.0067&0.0091&0.0075&\underline{0.0101}&0.0090&\textbf{0.0113}$\pm$0.0011&11.88\%\\
         &NDCG@10&0.0069&0.0083&0.0098&0.0124&0.0102&\underline{0.0132}&0.0121&\textbf{0.0148}$\pm$0.0008&12.12\%\\
         \midrule
         \multirow{4}*{Sports}&HR@5&0.0092&0.0137&0.0139&0.0218&0.0176&0.0157&\underline{0.0227}&\textbf{0.0326}$\pm$0.0007&43.61\%\\
         &HR@10&0.0188&0.0274&0.0231&0.0336&0.0326&0.0265&\underline{0.0374}&\textbf{0.0498}$\pm$0.0009&33.16\%\\
         &NDCG@5&0.0040&0.0096&0.0085&0.0127&0.0105&0.0098&\underline{0.0129}&\textbf{0.0208}$\pm$0.0010&61.24\%\\
         &NDCG@10&0.0051&0.0137&0.0126&0.0169&0.0153&0.0135&\underline{0.0184}&\textbf{0.0262}$\pm$0.0008&42.39\%\\
         \midrule
         \multirow{4}*{ML-1M}&HR@5&0.0078&0.0763&0.0816&0.1087&0.0733&0.1078&\underline{0.1147}&\textbf{0.2038}$\pm$0.0021&77.68\%\\
         &HR@10&0.0162&0.1658&0.1593&0.1904&0.1323&0.1952&\underline{0.1975}&\textbf{0.2946}$\pm$0.0018&49.16\%\\
         &NDCG@5&0.0052&0.0385&0.0372&0.0638&0.0432&0.0616&\underline{0.0662}&\textbf{0.1390}$\pm$0.0030&109.97\%\\
         &NDCG@10&0.0079&0.0671&0.0624&0.0910&0.0619&0.0917&\underline{0.0928}&\textbf{0.1680}$\pm$0.0032&81.03\%\\
         \midrule
         \multirow{4}*{Yelp}&HR@5&0.0127&0.0152&0.0156&0.0161&0.0186&0.0173&\underline{0.0216}&\textbf{0.0441}$\pm$0.0006&104.17\%\\
         &HR@10&0.0245&0.0263&0.0252&0.0265&0.0291&0.0282&\underline{0.0352}&\textbf{0.0631}$\pm$0.0010&79.26\%\\
         &NDCG@5&0.076&0.0104&0.0096&0.0102&0.0118&0.0114&\underline{0.0130}&\textbf{0.0325}$\pm$0.0004&150.00\%\\
         &NDCG@10&0.0119&0.0137&0.0129&0.0134&0.0171&0.0163&\underline{0.0185}&\textbf{0.0386}$\pm$0.0005&108.65\%\\
         \bottomrule
    \end{tabular}
    \label{tab:overall}
\end{table*}

\subsection{Setup}
\label{sec:setup}
\subsubsection{Dataset}
The experiments are conducted over five benchmark datasets with statistics after preprocessing shown in Table~\ref{tab:datasets}.
\begin{itemize}
    \item \textbf{Amazon Beauty}, \textbf{Clothing} and \textbf{Sports}~\cite{amazon}\footnote{http://jmcauley.ucsd.edu/data/amazon/}. Following baselines~\cite{sasrec,bert4rec,s3rec,cl4srec}, the widely used Amazon dataset is chosen in our experiments with three sub-categories.
    \item \textbf{MovieLens-1M}~\cite{movielens}\footnote{https://grouplens.org/datasets/movielens/1m/}. Following~\cite{bert4rec}, the popular movie recommendation dataset is used here, denoted as ML-1M.
    \item \textbf{Yelp}\footnote{https://www.yelp.com/dataset}, which is a widely used dataset for the business recommendation. Similar to~\cite{s3rec}, the transaction records after Jan. 1st, 2019 are used in our experiment.
\end{itemize}

Following~\cite{sasrec,bert4rec,s3rec,cl4srec} for preprocessing, all interactions are considered as implicit feedback. Users or items appearing less than five times are removed. The maximum length of a sequence is 50.

\subsubsection{Metrics}
\label{sec:metric}
For overall evaluation, top-$K$ Hit Ratio (HR@$K$) and top-$K$ Normalized Discounted Cumulative Gain (NDCG@$K$) are applied with $K$ chosen from $\{5,10\}$. We evaluate the ranking results over the whole item set for the fair comparison~\cite{metric}.

\subsubsection{Baselines}
\label{sec:baseline}
The following methods are used for comparison:
\begin{itemize}
    \item \textbf{BPR-MF}~\cite{bprmf} is the first method to use BPR loss to train a matrix factorization model.
    \item \textbf{GRU4Rec}~\cite{gru4rec} applies GRU to model the user sequence. It is the first recurrent model for sequential recommendation.
    \item \textbf{Caser}~\cite{caser} is a CNN-based method capturing high-order patterns by applying horizontal and vertical convolutional operations for sequential recommendation.
    \item \textbf{SASRec}~\cite{sasrec} is a single-directional self-attention model. It is a strong baseline in sequential recommendation.
    \item \textbf{BERT4Rec}~\cite{bert4rec} uses a masked item training scheme similar to the masked language model sequential in NLP. The backbone is the bi-directional self-attention mechanism.
    \item \textbf{$\text{S}^3\text{Rec}_\text{MIP}$}~\cite{s3rec} applied masked contrastive pre-training as well. The Mask Item Prediction (MIP) variant is used here.
    \item \textbf{CL4SRec}~\cite{cl4srec} uses item cropping, masking, and reordering as augmentations for contrastive learning. It is the most recent and strong baseline for sequential recommendation.
\end{itemize}

\subsubsection{Implementation}
\label{sec:imple}
The embedding size is set to $64$ with all linear mapping functions in DuoRec has the same hidden size. The numbers of layers and heads in the Transformer are set to $2$. The Dropout~\cite{dropout} rate on the embedding matrix and the Transformer module are chosen from $\{0.1,0.2,0.3,0.4,0.5\}$. The training batch size is set to $256$. We use the Adam~\cite{adam} optimizer with the learning $0.001$. $\lambda$ in Equation (\ref{eq:all-loss}) is chose from $\{0.1,0.2,0.3,0.4,0.5\}$.

\subsection{Overall Performance}
\label{sec:rq1}
In this experiment, we evaluate the overall performance to compare DuoRec with the baselines, which is presented in Table~\ref{tab:overall}.

According to the results, the first observation is that the non-sequential result, BPR-MF, can hardly achieve a comparable result with other sequential methods. When it comes to the deep learning era, the first representative method is GRU4Rec based on GRU, which can consistently outperform the non-sequential BPR-MF. It can be concluded that the incorporation of sequential information can improve performance. Similarly, Caser uses a convolutional module to aggregate the sequential tokens, which are stacked as a matrix. Caser generally has a similar performance to GRU4Rec. More recently, attention has become the strongest sequence encoder. SASRec is the first method to apply uni-directional attention for sequence encoding. Compared with the previous deep learning-based models, SASRec can improve the performance by a large margin. This is achieved by the more representative sequential encoder. More recent methods generally inherit the attention-based encoder while introducing extra objectives. For example, BERT4Rec applies the masked item prediction objective to enforce the model to understand the semantics by filling in the masks. Although such a task can introduce a meaningful signal for the model, the performance is not consistent since the masked item prediction is not aligned well with the recommendation task. A similar situation happens to $\text{S}^3\text{Rec}_\text{MIP}$, which also relies on the masked item prediction as the pre-training objective. The finetuning stage gives out a more accurate prediction. For the most recent contrastive learning-based approach, CL4SRec, achieves a consistent improvement over the other baselines. The extra objective is the same as the normal contrastive learning norm to set two different views of the same sequence. For DuoRec, it can outperform all the baselines by a large margin. The incorporation of the supervised and unsupervised positive samples can improve the overall performance by regularizing the distribution of the sequence and the item representation.

\begin{figure}[t]
    \centering
    \subfigure{
    \label{fig:nce-hr}
    \includegraphics[width=0.465\linewidth]{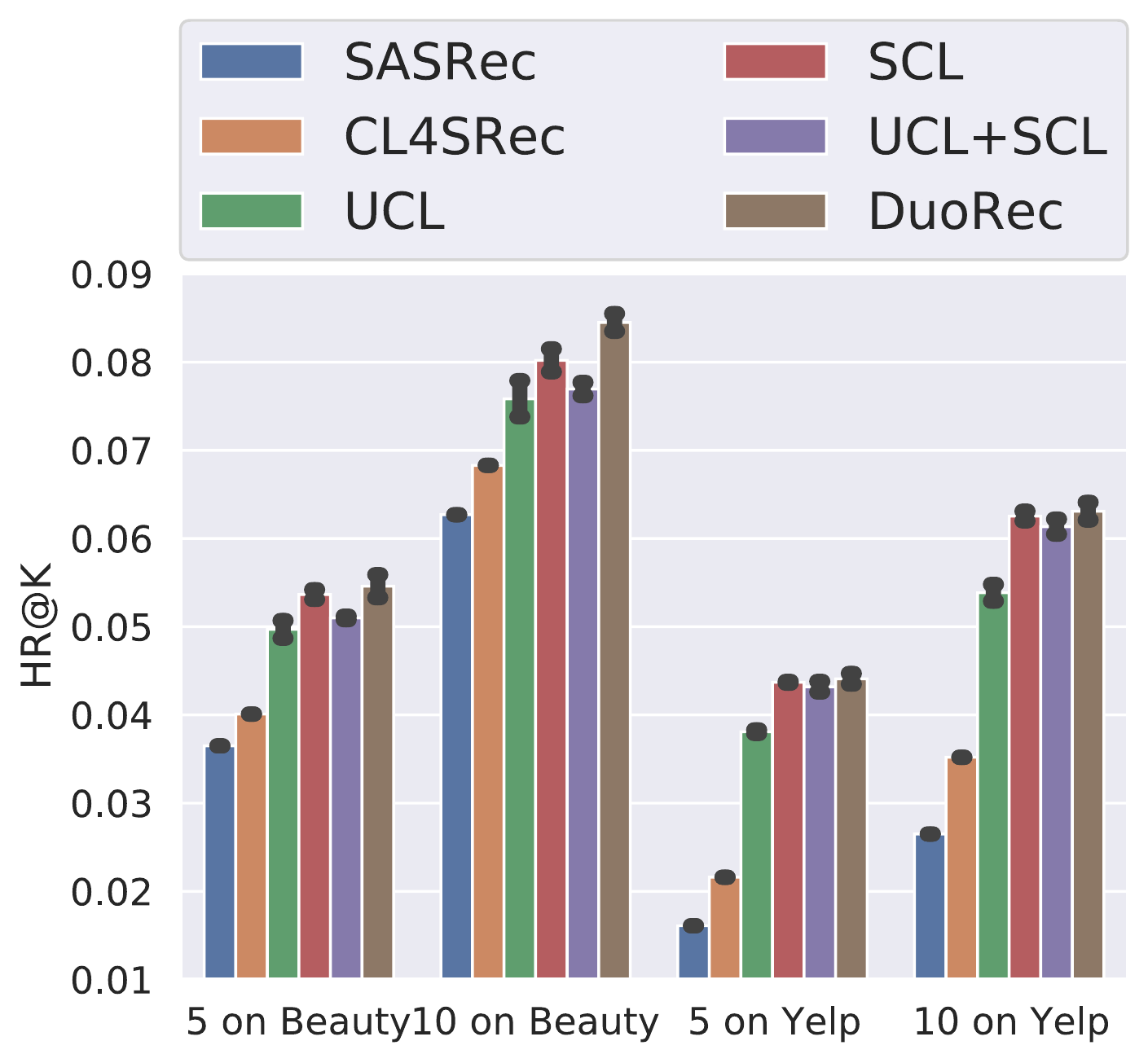}
    }
    \subfigure{
    \label{fig:nce-ndcg}
    \includegraphics[width=0.465\linewidth]{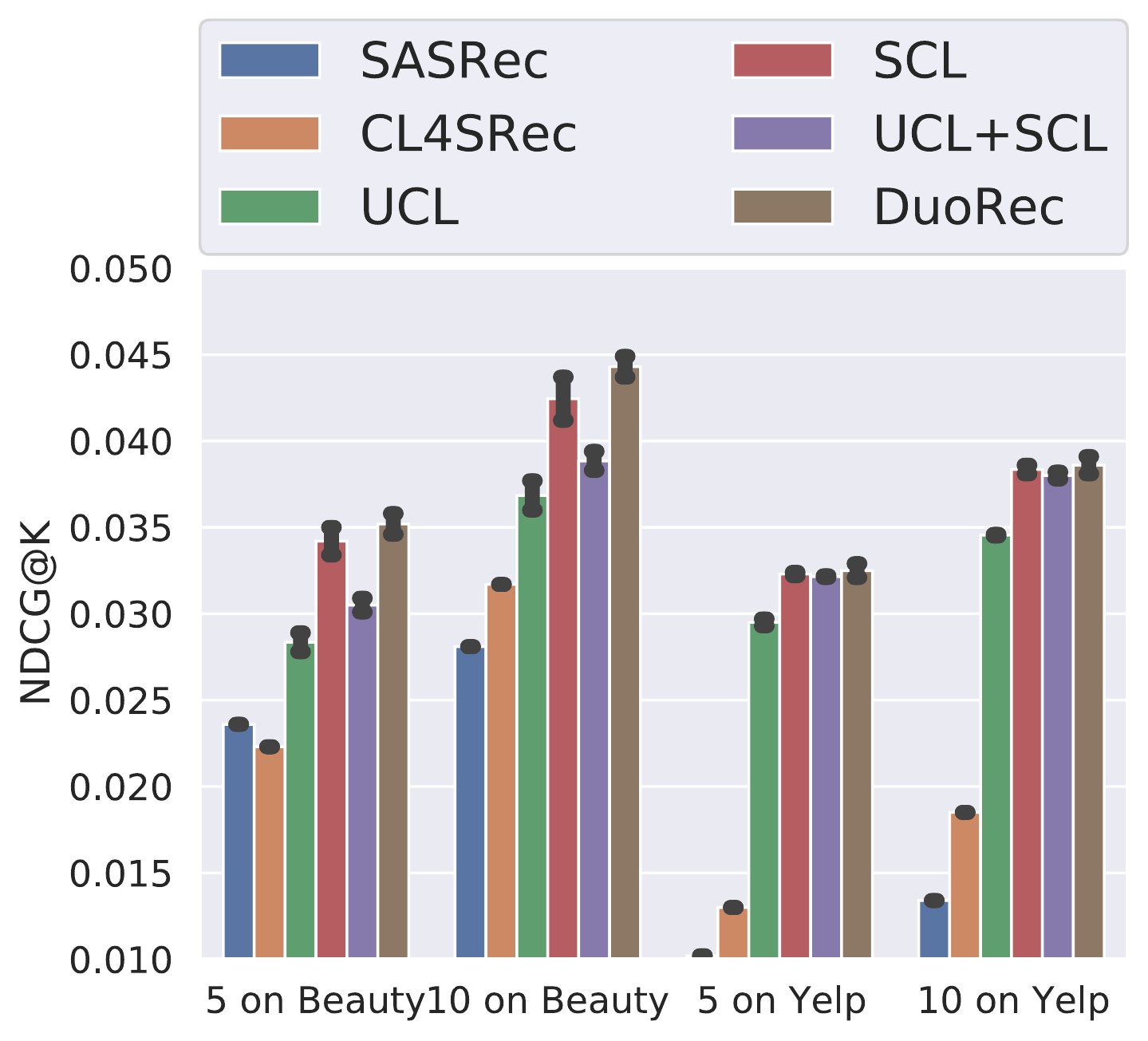}
    }
    \caption{Performance of different contrastive objectives.}
    \label{fig:nce}
\end{figure}

\begin{figure*}
    \centering
    \subfigure[SASRec.]{
    \label{fig:sas-clothing}
    \includegraphics[width=0.122\linewidth]{figure/SASRec-Amazon_Clothing_Shoes_and_Jewelry.png}
    }
    \subfigure[CL4SRec.]{
    \label{fig:cl-clothing}
    \includegraphics[width=0.123\linewidth]{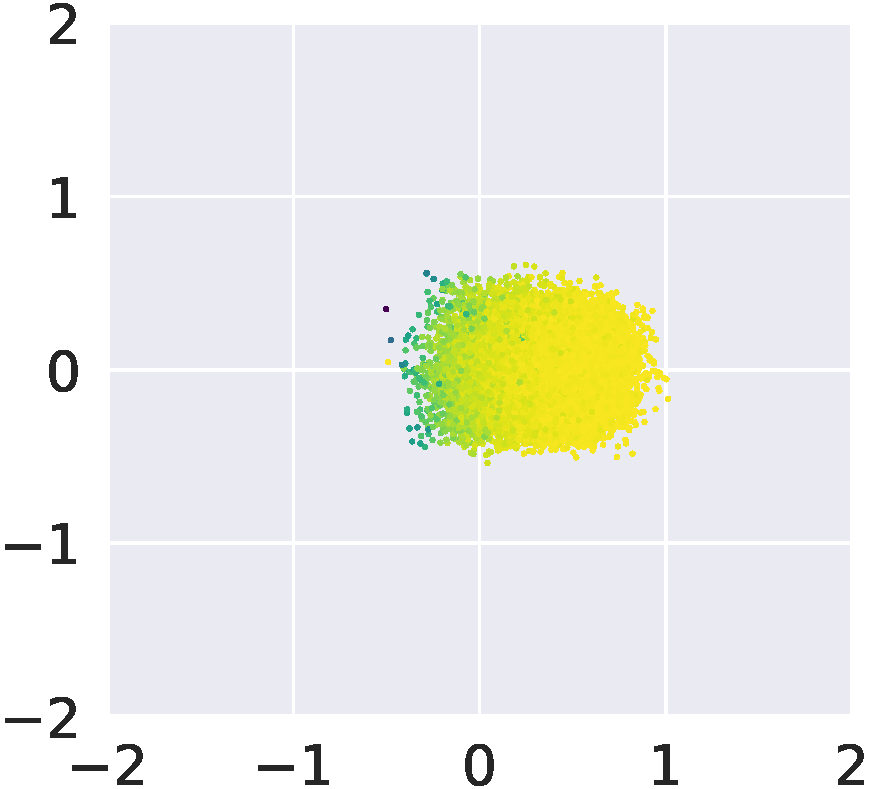}
    }
    \subfigure[UCL.]{
    \label{fig:ucl-clothing}
    \includegraphics[width=0.121\linewidth]{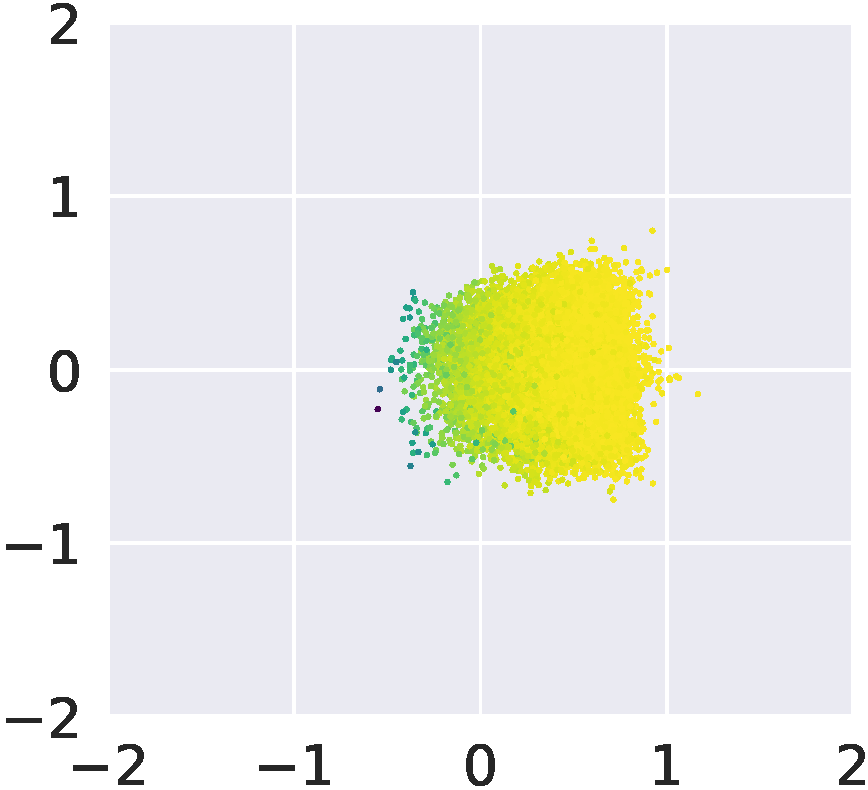}
    }
    \subfigure[SCL.]{
    \label{fig:scl-clothing}
    \includegraphics[width=0.121\linewidth]{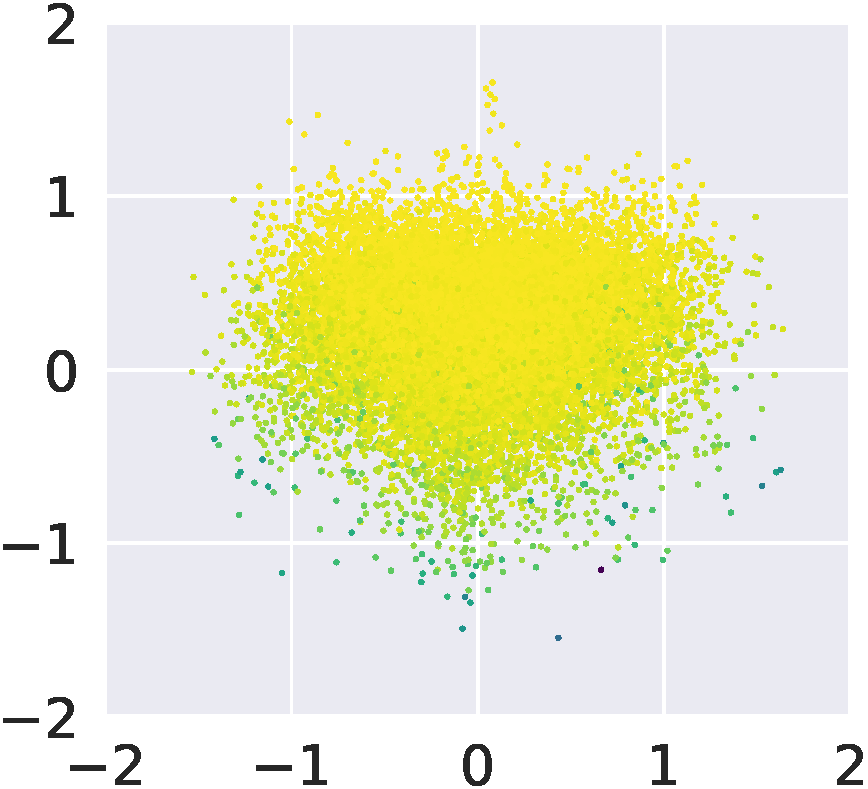}
    }
    \subfigure[UCL+SCL.]{
    \label{fig:us-clothing}
    \includegraphics[width=0.122\linewidth]{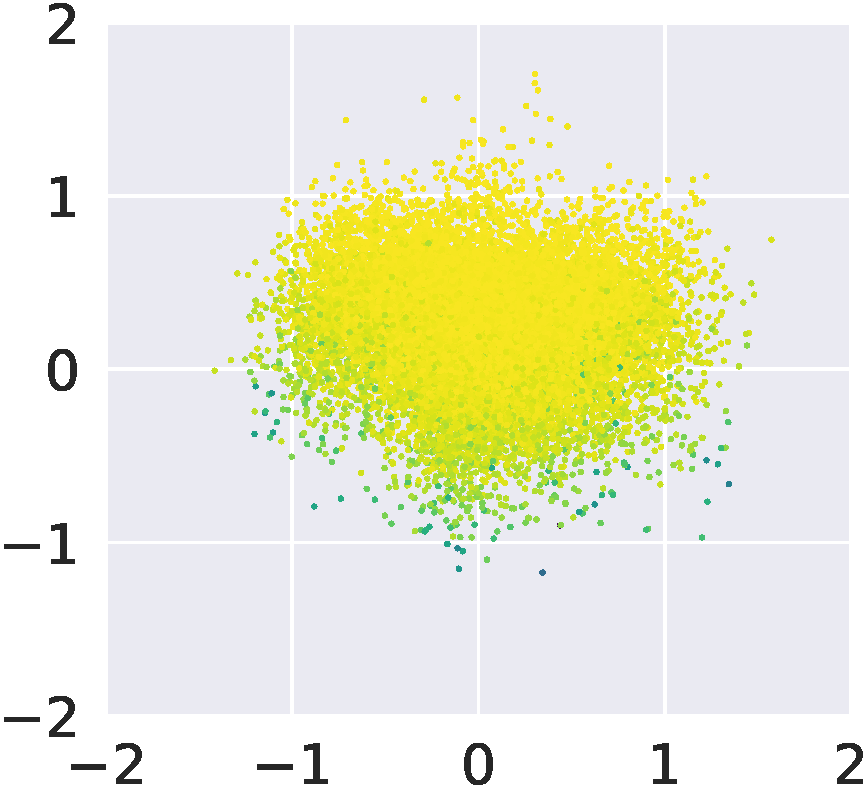}
    }
    \subfigure[DuoRec.]{
    \label{fig:USCLRec-clothing}
    \includegraphics[width=0.15\linewidth]{figure/USCLRec-Amazon_Clothing_Shoes_and_Jewelry.png}
    }
    \subfigure[Singular values.]{
    \label{fig:sv-clothing}
    \includegraphics[width=0.15\linewidth]{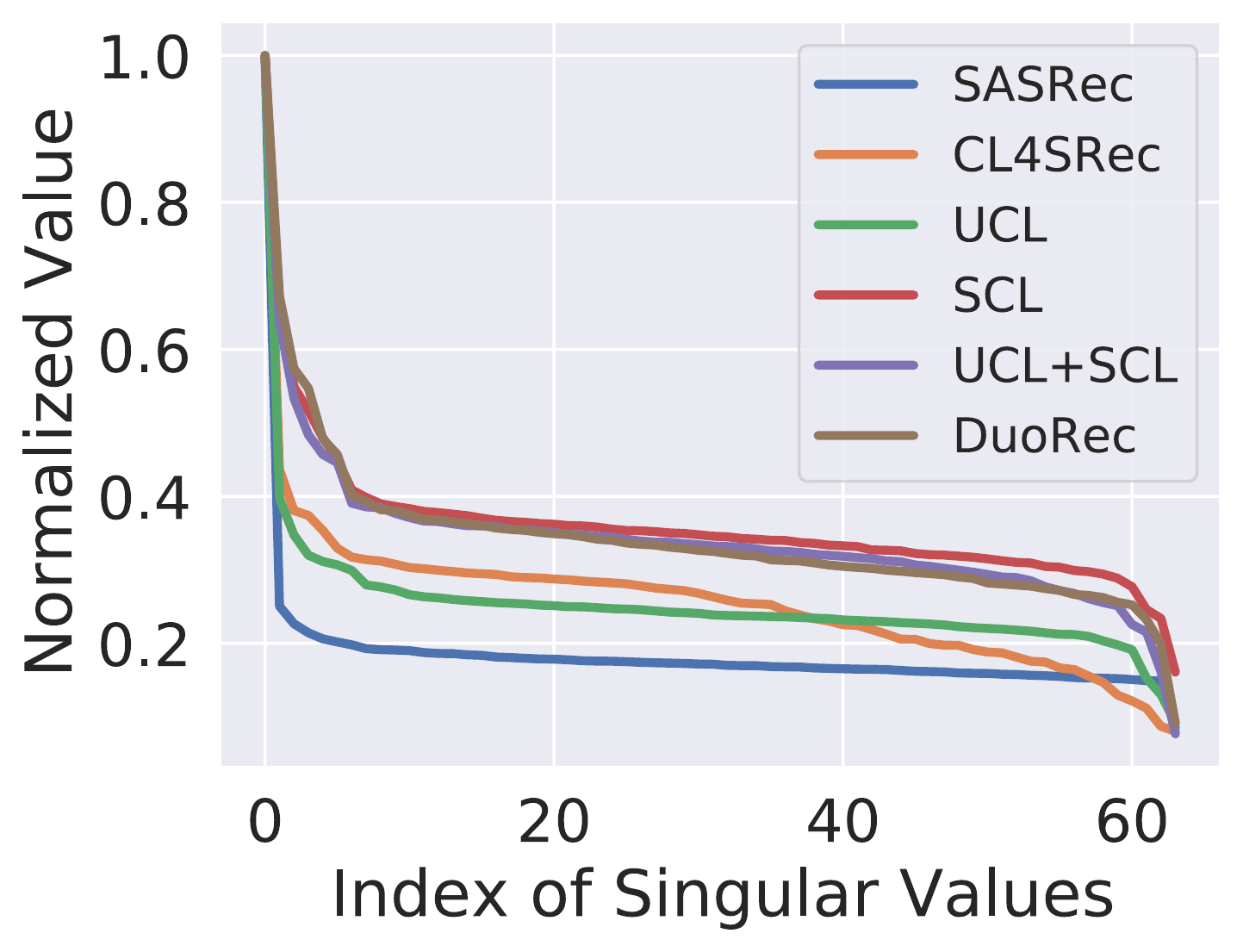}
    }
    \caption{Item embeddings on Clothing dataset.}
    \label{fig:svd-clothing}
\end{figure*}

\begin{figure*}
    \centering
    \subfigure[SASRec.]{
    \label{fig:sas-sports}
    \includegraphics[width=0.12\linewidth]{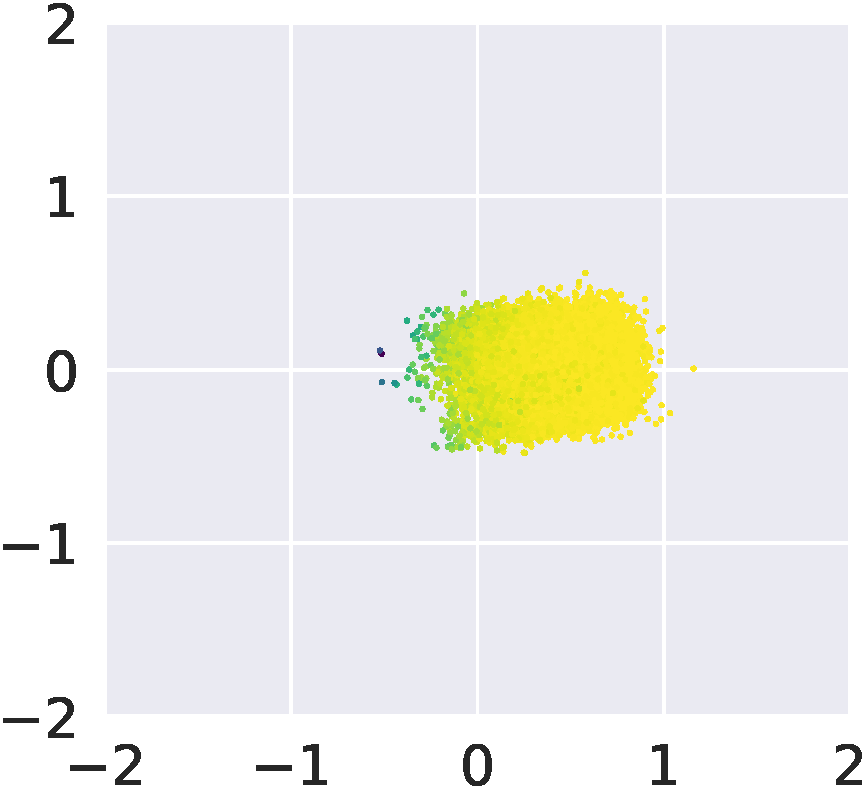}
    }
    \subfigure[CL4SRec.]{
    \label{fig:cl-sports}
    \includegraphics[width=0.121\linewidth]{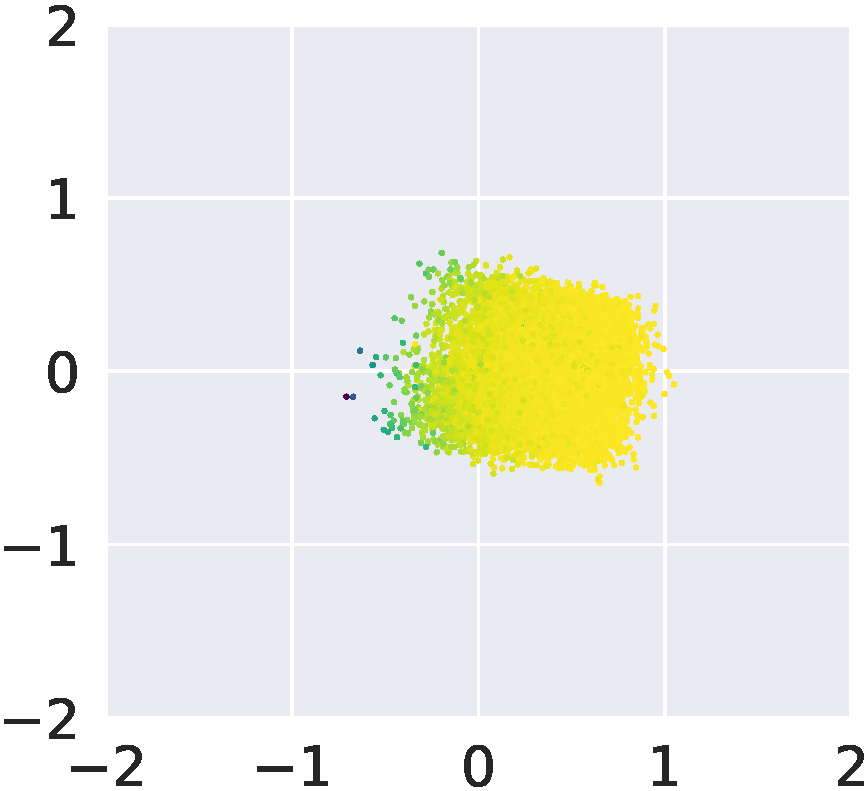}
    }
    \subfigure[UCL.]{
    \label{fig:ucl-sports}
    \includegraphics[width=0.122\linewidth]{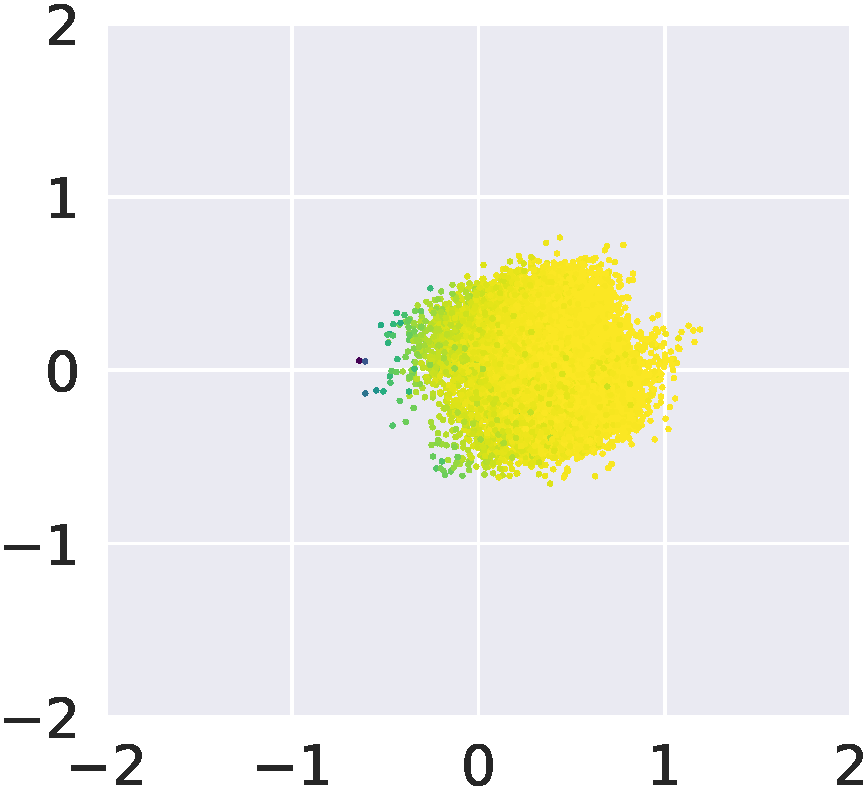}
    }
    \subfigure[SCL.]{
    \label{fig:scl-sports}
    \includegraphics[width=0.122\linewidth]{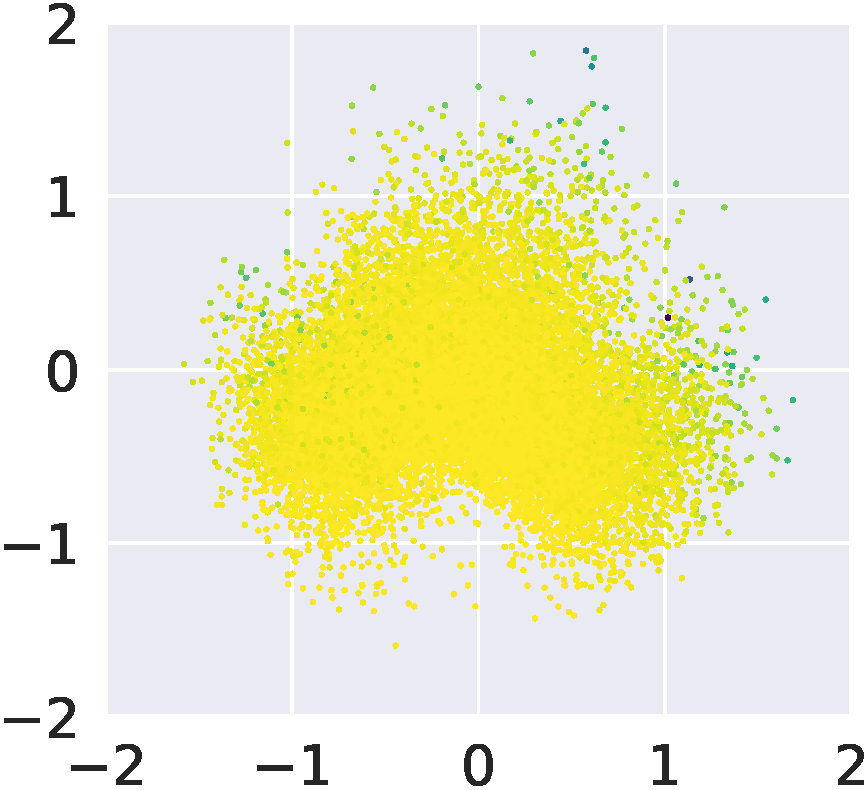}
    }
    \subfigure[UCL+SCL.]{
    \label{fig:us-sports}
    \includegraphics[width=0.122\linewidth]{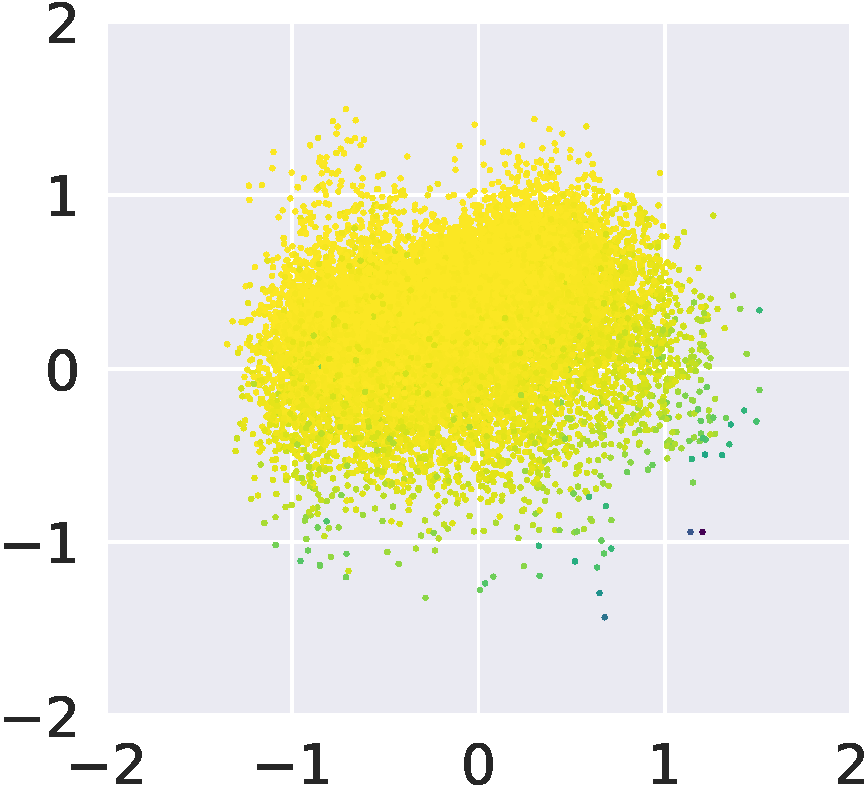}
    }
    \subfigure[DuoRec.]{
    \label{fig:USCLRec-sports}
    \includegraphics[width=0.156\linewidth]{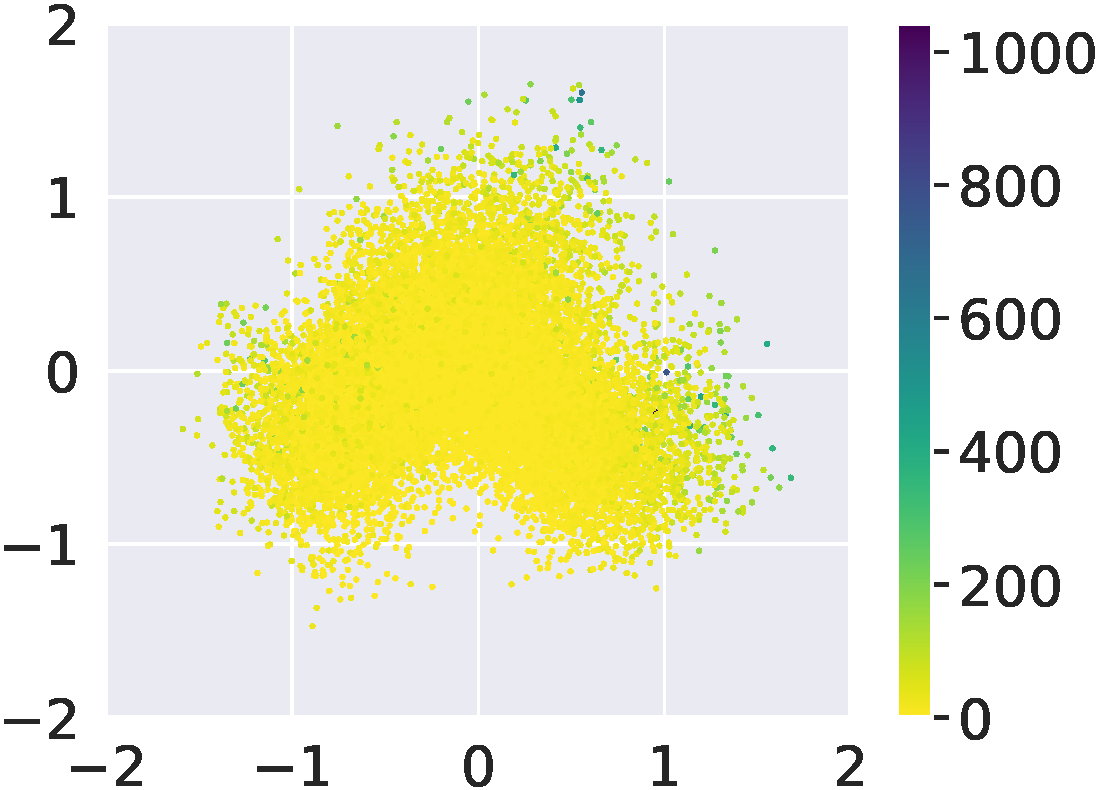}
    }
    \subfigure[Singular values.]{
    \label{fig:sv-sports}
    \includegraphics[width=0.15\linewidth]{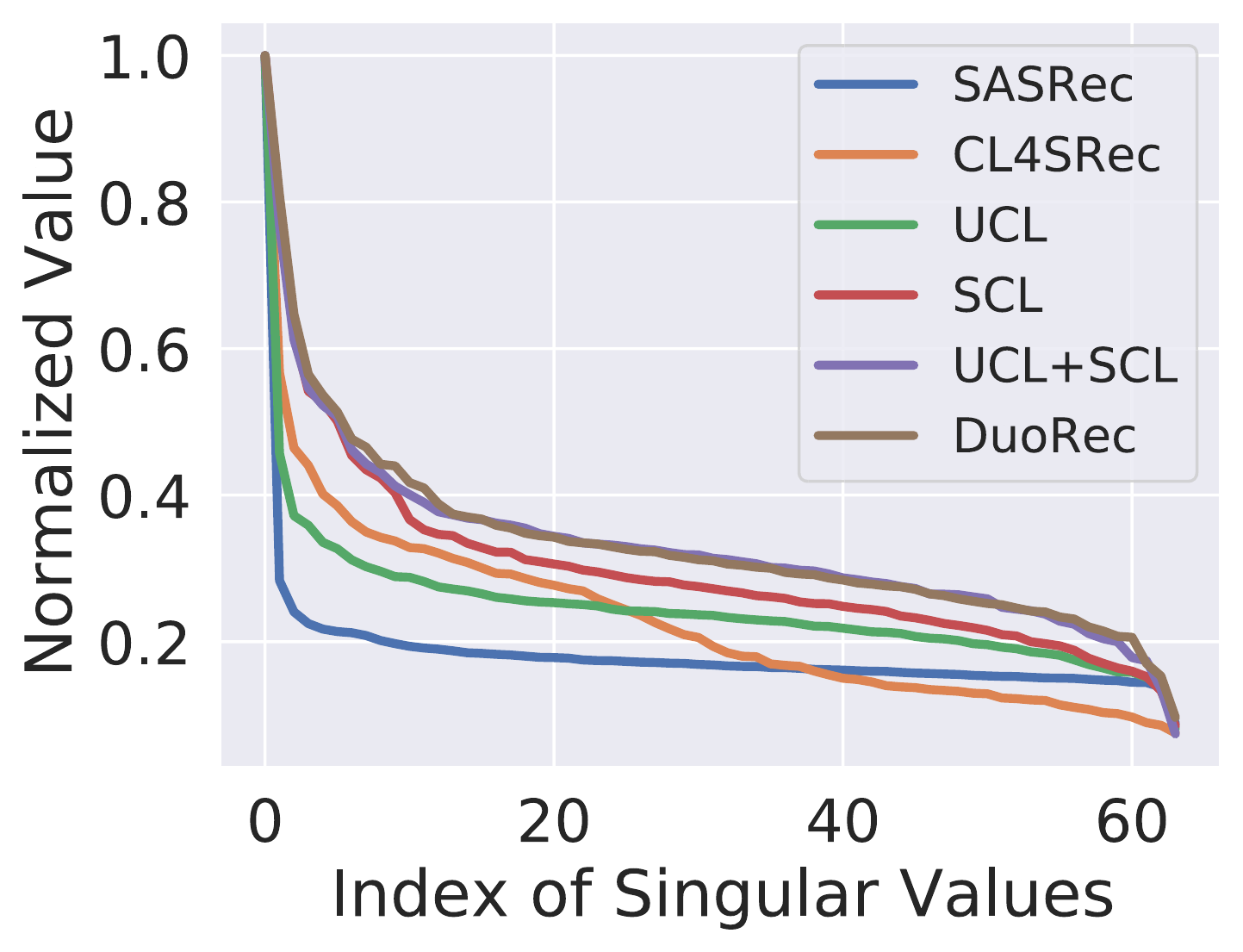}
    }
    \caption{Item embeddings on Sports dataset.}
    \label{fig:svd-sports}
\end{figure*}

\subsection{Ablation Study of Contrastive Learning}
\label{sec:rq-cl}
In this experiment, the efficacy of the unsupervised augmentation and the supervised positive sampling is evaluated. The variants are: CL4SRec, using cropping, masking, and reordering as augmentations to calculate NCE; UCL, the NCE uses unsupervised augmentations to optimize; SCL, the NCE uses supervised positive sampling to optimize; and UCL+SCL, trained with the addition of the UCL and the SCL losses. The result is shown in Figure~\ref{fig:nce}.

From the result, it is clear that adding a contrastive objective can generally improve the recommendation performance compared with the baseline SASRec. Compared with CL4SRec, UCL can outperform CL4SRec while both being unsupervised contrastive methods. It can be concluded that the model-level Dropout augmentation can provide a more semantically consistent unsupervised sample than the data-level augmentation. Furthermore, SCL relies on the target item to sample a semantically consistent supervised sample, which shows a large margin improvement over both the unsupervised methods. Interestingly, directly adding the UCL and the SCL losses will harm the performance. This could be due to the incompatible alignment of two contrastive losses. For DuoRec, both the unsupervised and supervised positive samples are exploited, which can yield the best performance compared with all other methods.

\begin{figure}[t]
    \centering
    \subfigure[Clothing.]{
    \label{fig:cloth-loss}
    \includegraphics[width=0.46\linewidth]{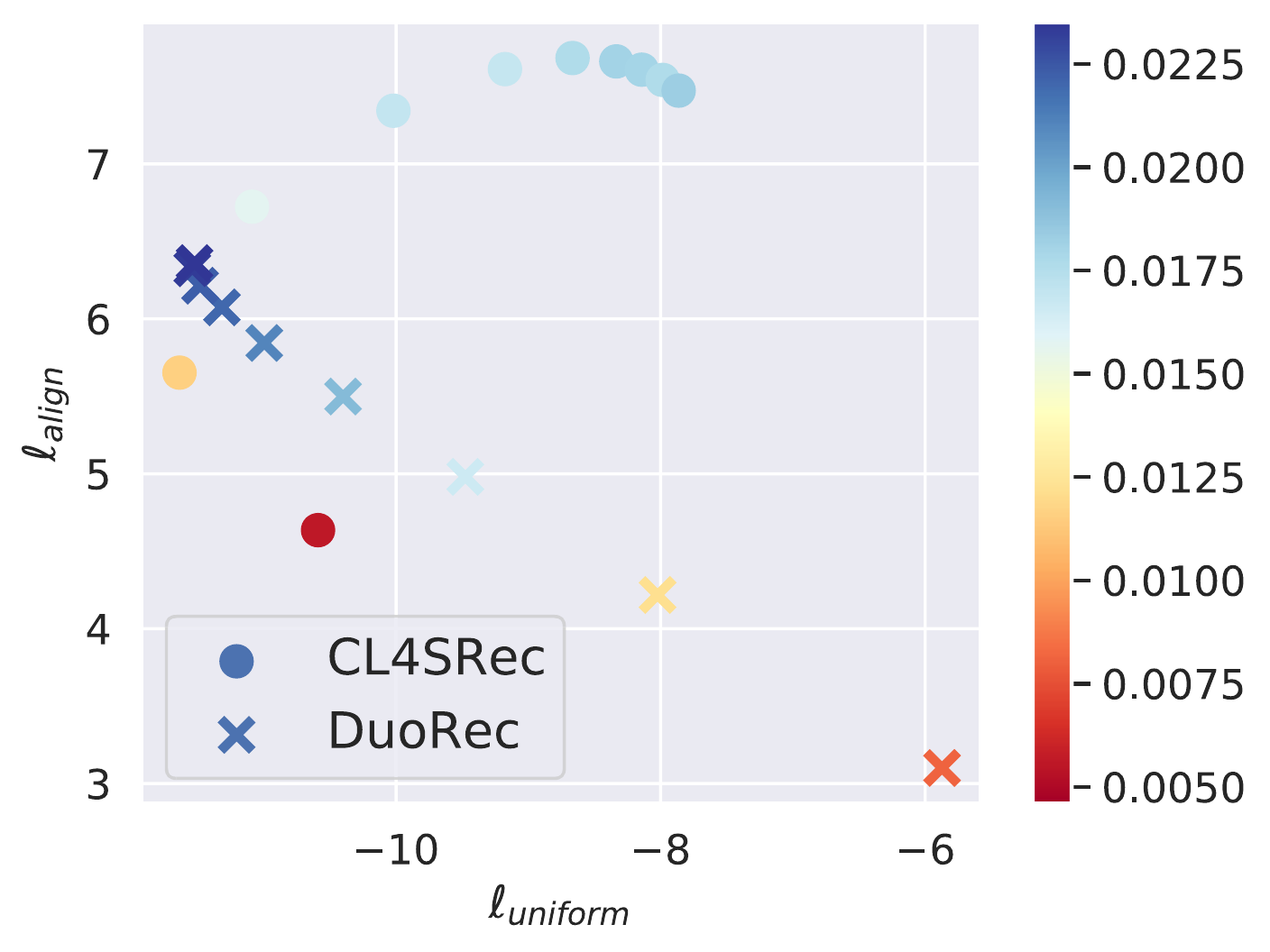}
    }
    \subfigure[Yelp.]{
    \label{fig:yelp-loss}
    \includegraphics[width=0.47\linewidth]{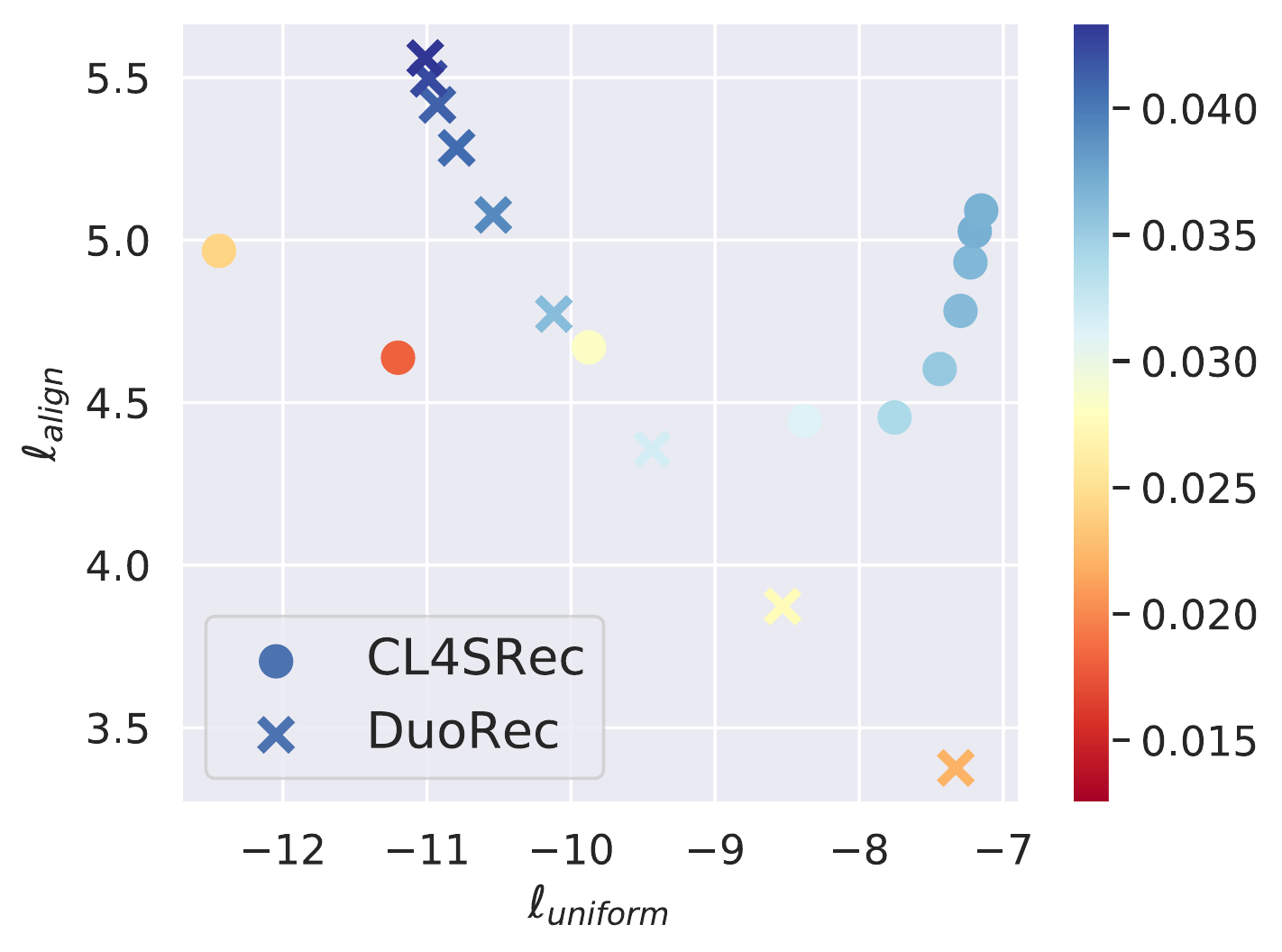}
    }
    \caption{Training loss with colors indicating the validation HR@5 (blue is better). The uniformity loss of sequence representations decreases during training of DuoRec as the HR@5 increases, which indicates a more uniform distribution. In contrast, the uniformity loss increases during training of CL4SRec, resulting in an anisotropic distribution.}
    \label{fig:loss}
\end{figure}

\subsection{Contrastive Regularization in Training}
\label{sec:rq-vis-cl}
To evaluate how the contrastive regularization affects the training, (1) the visualization of the learned embedding matrix and (2) the training losses, will be demonstrated to help understand how contrastive learning improves performance. The visualization is based on SVD decomposition, which will project the embedding matrix into 2D and give out the normalized singular values. The results are shown in Figure~\ref{fig:svd-clothing} and~\ref{fig:svd-sports}. The visualization of the training losses is decomposed into the alignment and the uniformity via Equation (\ref{eq:align}) and (\ref{eq:uniform}) as presented in Figure~\ref{fig:loss}.

\subsubsection{Visualization of Item Embedding}
As discussed before, SASRec is trained without constraint on the embedding matrix, which yields a narrow cone in the latent space as in Figure~\ref{fig:sas-clothing} and~\ref{fig:sas-sports}. The resulted in singular values drastically decrease to very small values. Although CL4SRec has an extra contrastive loss with data-level augmentation, the embeddings improve in terms of the distribution magnitude while the rare items are still located on the same side of the origin point. And the singular values of CL4SRec decrease slower than SASRec. This could be due to the data-level augmentation cannot consistently provide reasonable sequences. While for the UCL variant, it generates a similar embedding distribution as CL4SRec since they are both based on unsupervised contrastive learning. While for SCL, which only uses supervised contrastive learning, it is clear that the distribution of embeddings is more balanced that both the high- and low-frequency items are located around the origin point. The singular values are significantly higher than the unsupervised methods. It can be concluded that the supervised positive samples are more semantically consistent with the input sequence. When adding both the unsupervised and the supervised positive samples, UCL+SCL has a similar situation as purely SCL for the Clothing dataset while different for the Sports dataset. This difference is due to the combination of the unsupervised and supervised contrastive losses, which could lead the model in a different training direction. For the DuoRec, the embedding is distributed in a balanced style with the singular values decrease slowly. The proper combination of unsupervised and supervised contrastive learning improves the item embedding distribution.

\subsubsection{Alignment and Uniformity}
To investigate how contrastive learning takes effect during the training, the alignment loss term and the uniformity loss term are illustrated (both are better when smaller). The uniformity is calculated within the original output sequence representation, which is in the same range of samples for every method. Noting that since the choices of the positive sample are different across different methods, the alignment term is shown as a trend indicator without a proper meaning for comparison.

From Figure~\ref{fig:cloth-loss} and~\ref{fig:yelp-loss}, it is clear that as the training of DuoRec goes on, the uniformity loss decreases as the HR@5 increases. And the uniformity loss achieves a clearly lower value than CL4SRec, which reflects the sequence representations are distributed more uniformly. For the CL4SRec, the uniformity loss increases during the training, which indicates a worse distribution space compared with DuoRec. Although the alignment loss is slightly increasing during the training of both methods, the drop of uniformity loss actually improves the recommendation performance.

\begin{figure}[t]
    \centering
    \subfigure{
    \label{fig:dropout-hr5}
    \includegraphics[width=0.465\linewidth]{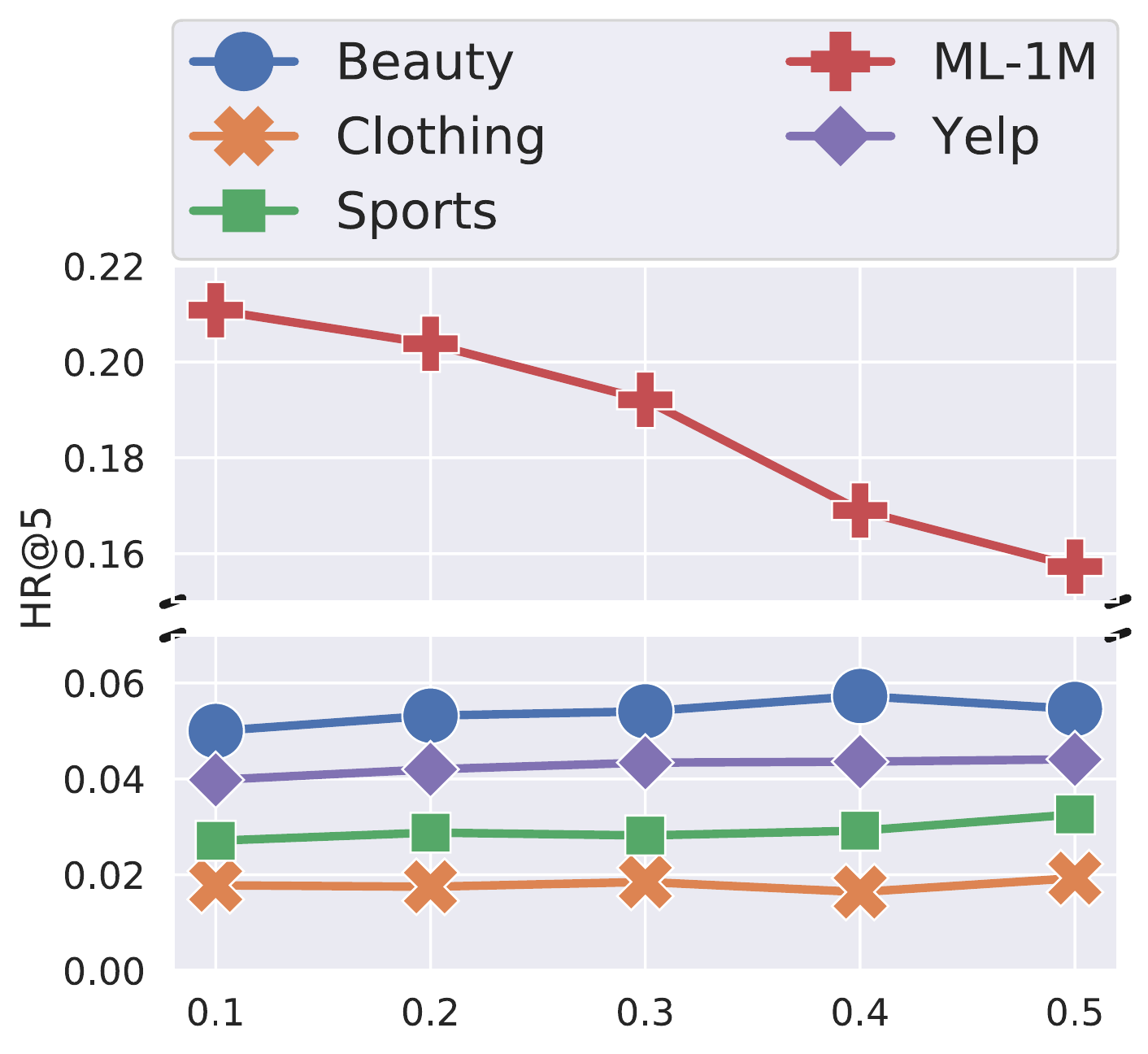}
    }
    \subfigure{
    \label{fig:dropout-hr10-hr}
    \includegraphics[width=0.465\linewidth]{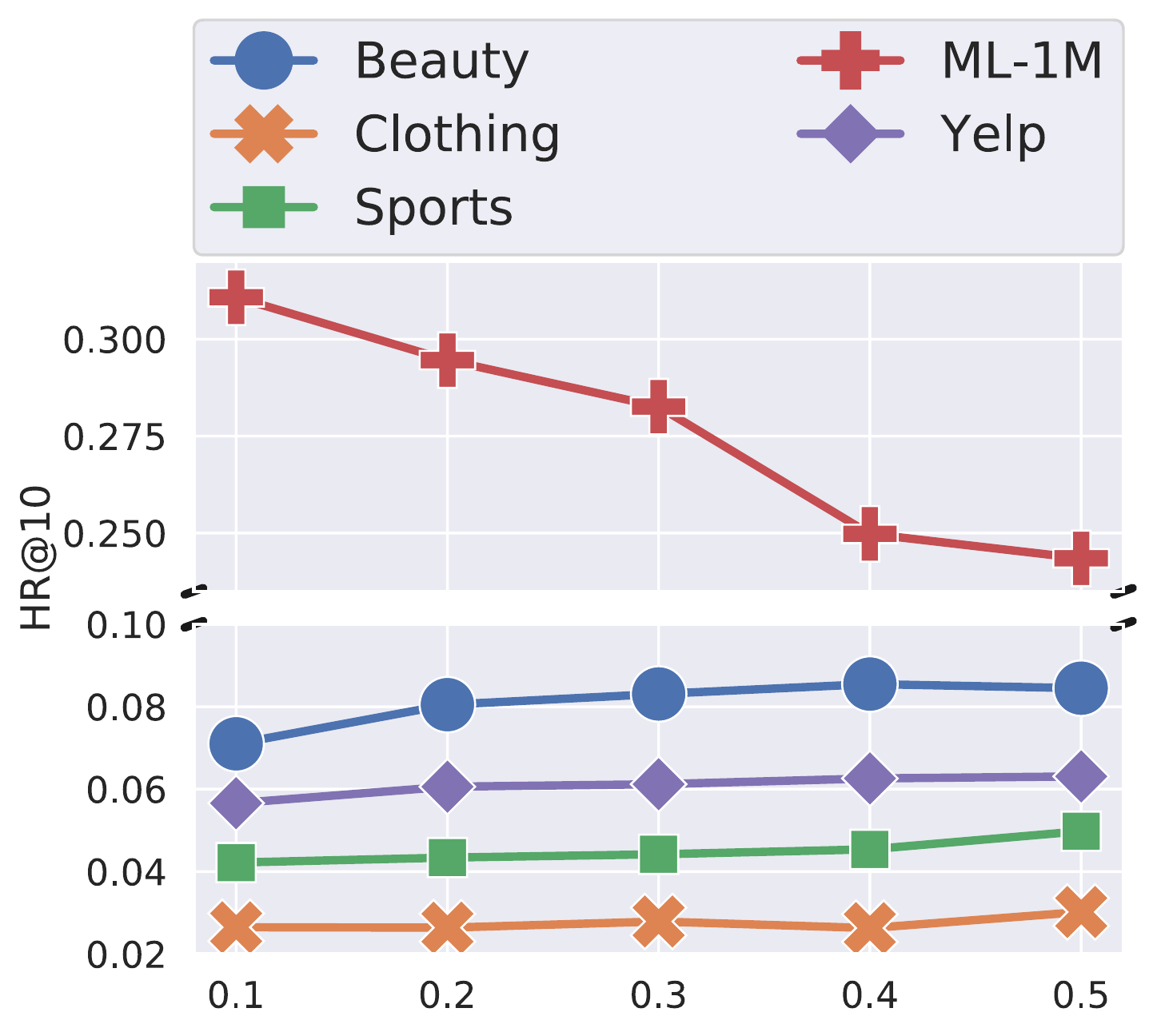}
    }
    \caption{Parameter sensitivity of Dropout ratio.}
    \label{fig:param-dropout}
\end{figure}

\begin{figure}[t]
    \centering
    \subfigure{
    \label{fig:lambda-hr5}
    \includegraphics[width=0.465\linewidth]{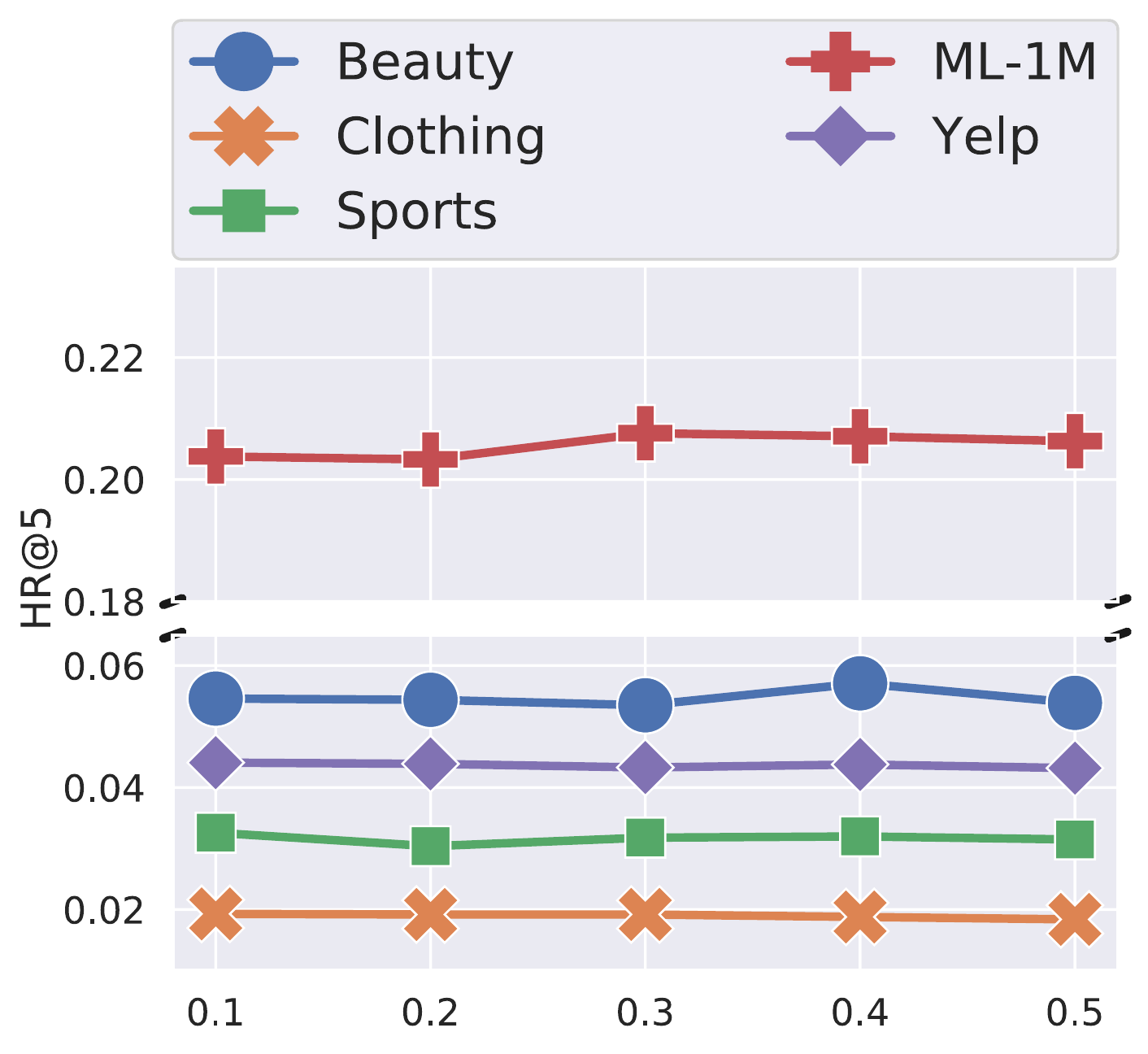}
    }
    \subfigure{
    \label{fig:lambda-hr10}
    \includegraphics[width=0.465\linewidth]{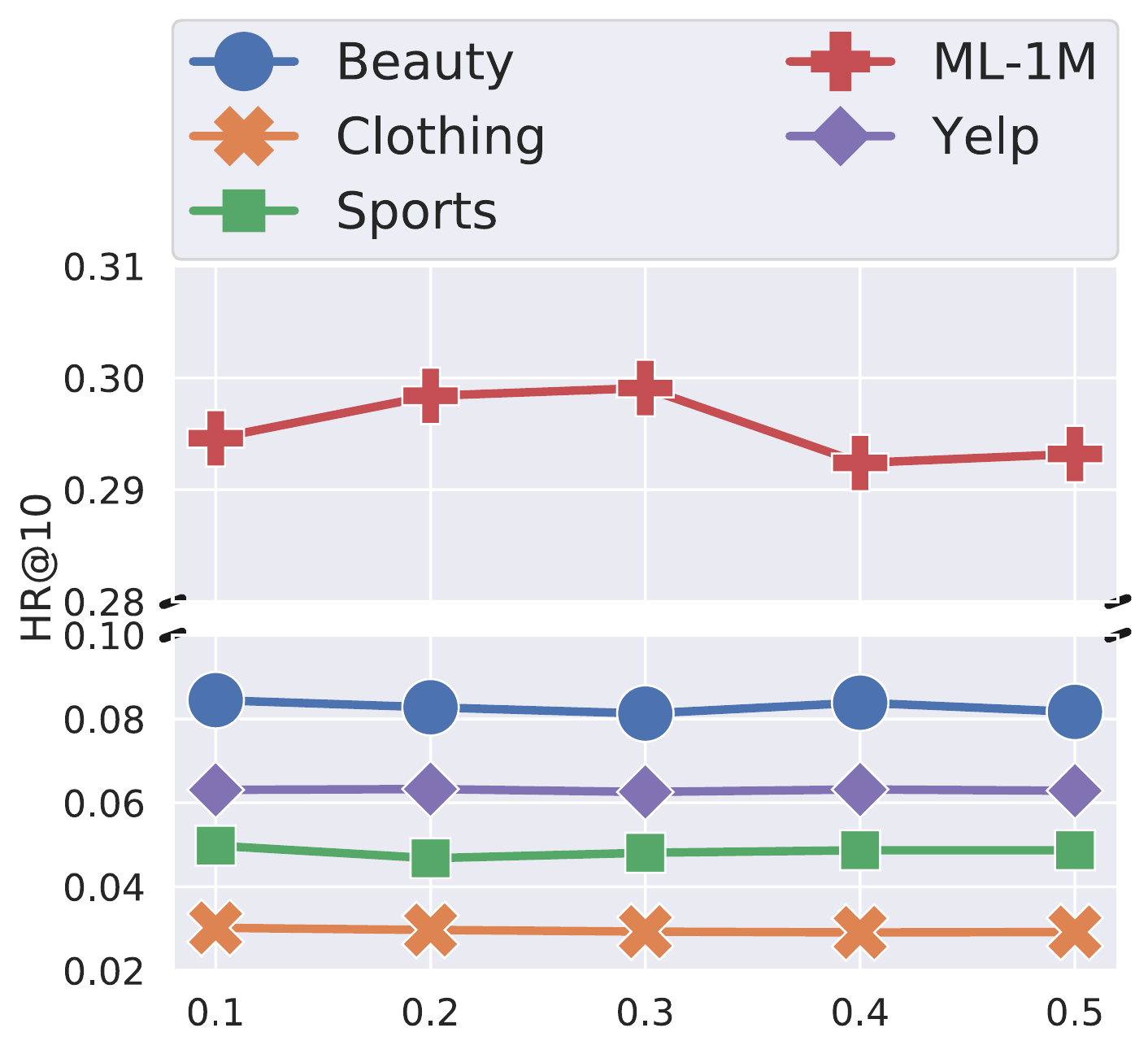}
    }
    \caption{Parameter sensitivity of $\lambda$.}
    \label{fig:param-lambda}
\end{figure}

\subsection{Parameter Sensitivity}
\label{sec:rq-param}
In this experiment, the parameter sensitivity of the Dropout ratio for the augmentation and the $\lambda$ in Equation (\ref{eq:all-loss}) are investigated. The results are presented in Figure~\ref{fig:param-dropout} and~\ref{fig:param-lambda} respectively. More results for reproducibility can be found in the appendices.

The Dropout ratio mainly affects the unsupervised augmentation, which assumes that using different Dropout masks under the same weight can generate semantically similar samples. According to Figure~\ref{fig:param-dropout}, for ML-1M dataset, when increasing the Dropout ratio, the performance decreases, which could be due to the density for this dataset is higher with more training signals. When the augmentation is too different from the original input using a higher Dropout ratio, the model is guided to train in an inaccurate direction. While for the Beauty, Clothing, Sports and Yelp datasets, the effect of different Dropout ratios is not significant.

The $\lambda$ in Equation (\ref{eq:all-loss}) controls the scale of the contrastive regularization. According to Figure~\ref{fig:param-lambda}, the performance is consistent across different choices of $\lambda$. This is possibly because the contrastive regularization is well aligned with the recommendation task.

\section{Related Work}
\label{sec:rl}
\subsection{Sequential Recommendation}
The sequential recommendation task is mainly related to the sequential modeling methods~\cite{gru4rec,caser,sasrec,bert4rec,s3rec,s2s,safm,cl4srec,consensus1,consensus2,discrete}, which rely on recurrent neural networks such as GRU~\cite{gru} or attention structures~\cite{attention} as the sequence encoder. GRU4Rec~\cite{gru4rec} is the very first attempt to utilize the GRU network in sequential recommendation. Since the attention mechanism has shown a great ability, different related models are developed, e.g., SASRec~\cite{sasrec}, BERT4Rec~\cite{bert4rec} and $\text{S}^3$Rec~\cite{s3rec}. Recent graph-based methods, FGNN~\cite{fgnn,fgnnj}, GAG~\cite{gag} and PosRec~\cite{posrec}, achieve improved performance due to the graph modeling in sequence. In terms of the training method, most methods are based on the next-item prediction task~\cite{gru4rec,caser,sasrec}. These methods are naturally suitable for the next-item prediction problem. The other training scheme usually has extra training tasks~\cite{bert4rec,s3rec,cl4srec,s2s}. The pre-training tasks mainly have the masked item prediction, the masked attribute or segment prediction in $\text{S}^3$Rec~\cite{s3rec} and the finetuning step is the same as the next-item prediction. A recent work MMInfoRec~\cite{mminforec} applies an item level contrastive learning for feature-based sequential recommendation. For BERT4Rec~\cite{bert4rec} and CL4SRec~\cite{cl4srec}, the auxiliary task is added in the multi-tasking style.

\subsection{Contrastive Learning}
Contrastive learning has been widely used in various deep learning areas for its strong ability to help with the self-supervised learning~\cite{moco,simclr,cpc,ecpc,deepinfomax,memdpc,coclr,simcse,simsiam,ldsdg}. For the computer vision problems, the early method such as CPC~\cite{cpc,ecpc} and DIM~\cite{deepinfomax}, the encodings of different scales of the same image are fed into the contrastive learning as positive pairs. In the follow-up methods e.g., MoCo~\cite{moco}, SimCLR~\cite{simclr} and SimSiam~\cite{simsiam}, the different augmentations of the same image are considered as positive pairs for the contrastive learning. For the video representation learning, MemDPC~\cite{memdpc} applies a similar strategy as CPC and DIM to encode the feature vectors of the segment and the video clip as positive pairs. For the COCLR~\cite{coclr} method, the positive samples of a video clip in RGB space are determined by the closeness of clips in the optical flow space and vice versa. In contrastive learning in the language modeling, ConSERT~\cite{consert} introduces traditional augmentation methods such as cropping and reordering into the sentence augmentation as positive pairs. SimCSE~\cite{simcse} treats the same sentence with different Dropout~\cite{dropout} masks as positive pairs.

Contrastive learning is used in recent recommendation methods. For the collaborative filtering methods, SGL~\cite{sgl} applies the NCE in node-level representation learning. SSL~\cite{simclrec} proposes a siamese network to encode the items as pre-training with embedding-level augmentations. SEPT~\cite{sstt} applies the NCE for the socially-aware recommendation, which is based on node-level contrastive learning. For the contrastive learning in sequential recommendation, $\text{S}^3$Rec~\cite{s3rec} incorporates the contrasting mechanism between the prediction and the ground truth of the attribute-level, the item-level, and the segment-level together for training. The segment-level contrastive learning is also applied by Ma et al.~\cite{s2s} as a multi-tasking objective. CL4SRec~\cite{cl4srec} proposes three augmentations for the interaction sequence and applies a similar contrastive strategy as MoCo~\cite{moco} and SimCLR~\cite{simclr} to set these augmentations of the same sequence as the positive pair in training. DHCN~\cite{sslsb} and MHCN~\cite{channel} are graph-based methods with contrastive learning on node-level representation. A more recent work MMInfoRec~\cite{mminforec} has achieved a great improvement in sequential recommendation with side information using a contrastive objective in item level via Dropout augmentation in the item embedding.

\section{Conclusion}
\label{sec:conclusion}
In this paper, the representation degeneration problem of the item embedding matrix in sequential recommendation is investigated. The empirical observation and the theoretical analysis are provided. To solve this problem, a DuoRec model is proposed, which contains a contrastive regularization with both the Dropout-based model-level augmentation and the supervised positive sampling to construct contrastive samples. The properties of this regularization term are analyzed towards the representation degeneration problem. Extensive experiments are conducted on five benchmark datasets, which verify the superiority of DuoRec. The visualization is demonstrated to show how DuoRec solves this problem.

\section{Acknowledgments}
The work is supported by Australian Research Council (CE200100025, DP190102353, DP190101985, FT210100624).

\clearpage
\bibliographystyle{ACM-Reference-Format}
\bibliography{acmart}

\clearpage
\appendix
\section{Results of Different Batch Sizes}
\begin{figure}[!h]
    \centering
    \subfigure{
    \label{fig:batch-hr5}
    \includegraphics[width=0.465\linewidth]{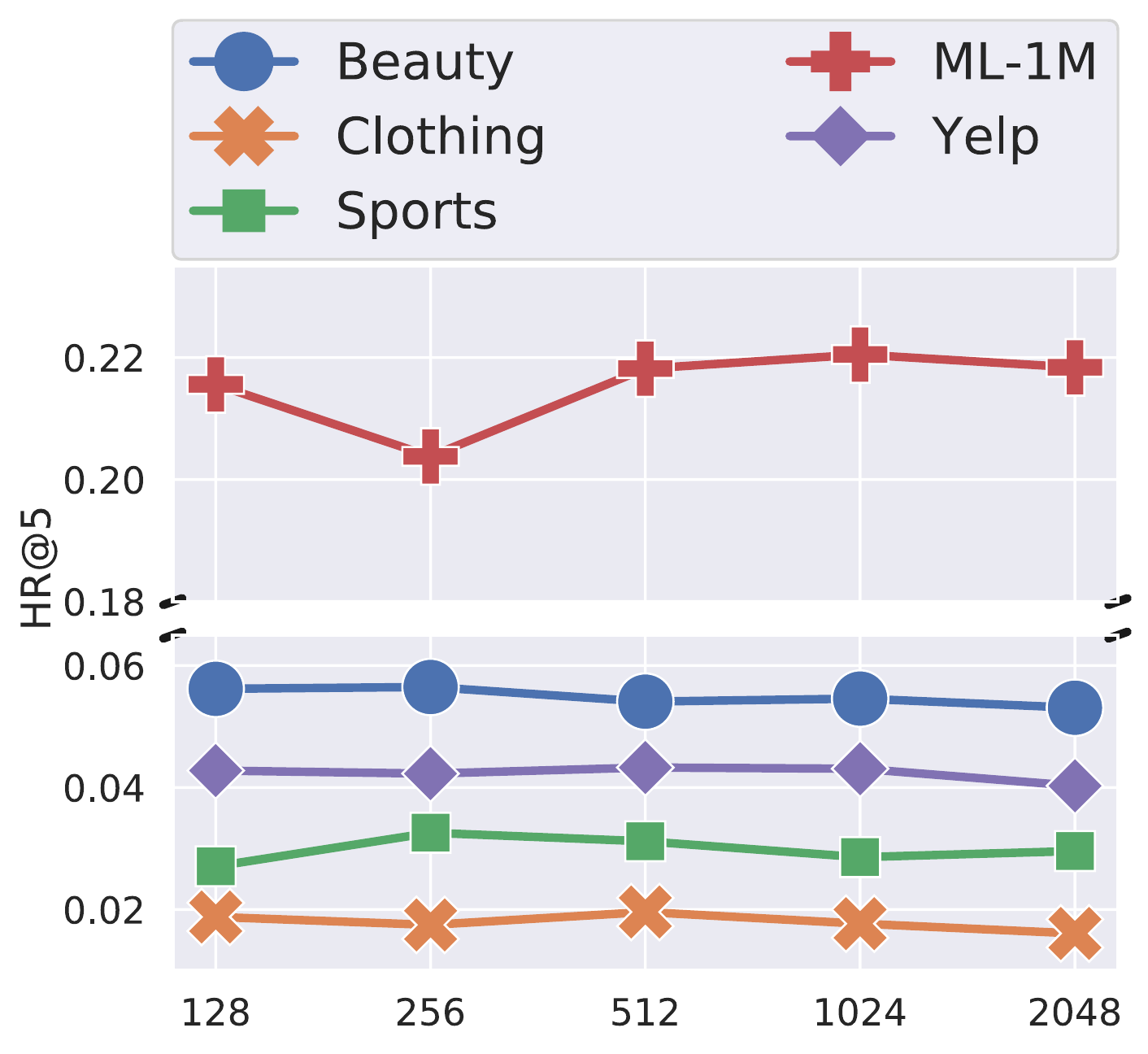}
    }
    \subfigure{
    \label{fig:batch-hr10}
    \includegraphics[width=0.465\linewidth]{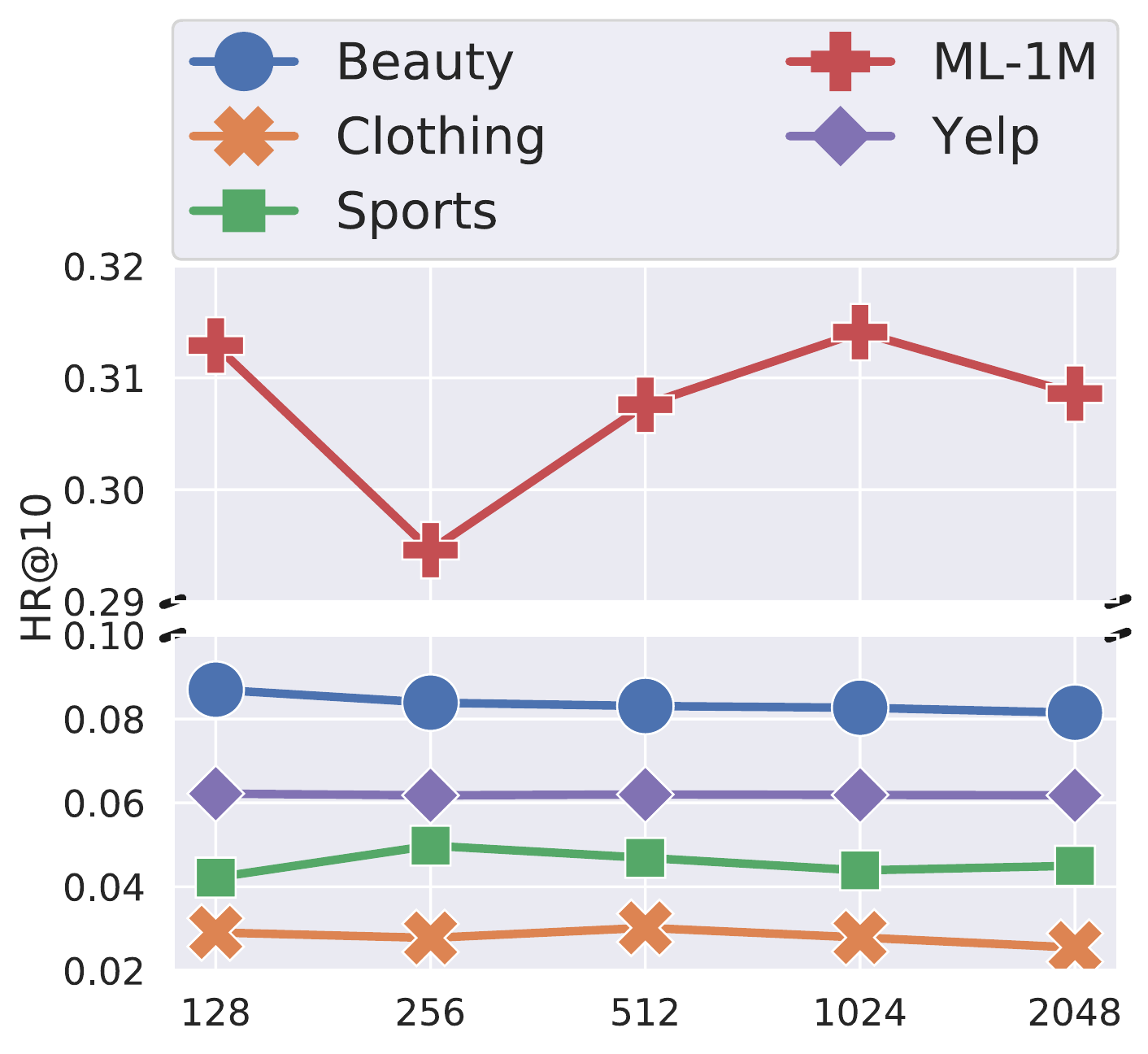}
    }
    \caption{Parameter sensitivity of batch size.}
    \label{fig:param-batch}
\end{figure}

For contrastive learning approaches, the sample size is an important hyper-parameter of the model since the sample size can affect the estimation of mutual information via the approximation of NCE~\cite{vbmi}. In Figure~\ref{fig:param-batch}, the experimental results of different batch sizes are presented. Under the choice from $\{128, 256, 512, 1024, 2048\}$, the performance has a slight decrease but in an acceptable range. This could be due to the multi-tasking learning paradigm, where the batch size affects not only the contrastive learning, but also the recommendation task.

\section{Results of Different Dropout Strategies}
\begin{figure}[!h]
    \centering
    \subfigure[HR@5.]{
    \label{fig:beauty-dp-hr5}
    \includegraphics[width=0.465\linewidth]{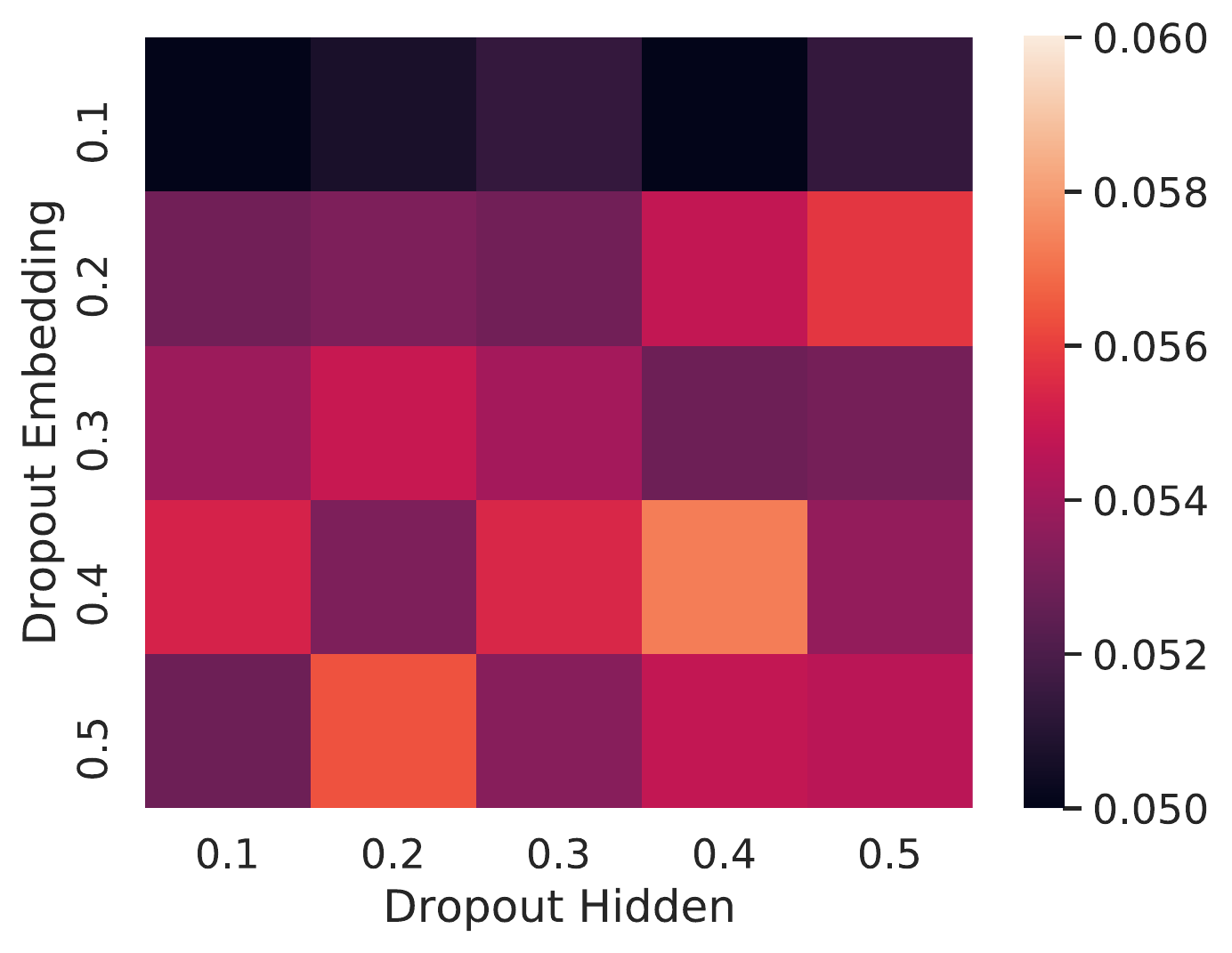}
    }
    \subfigure[HR@10.]{
    \label{fig:beauty-dp-hr10}
    \includegraphics[width=0.465\linewidth]{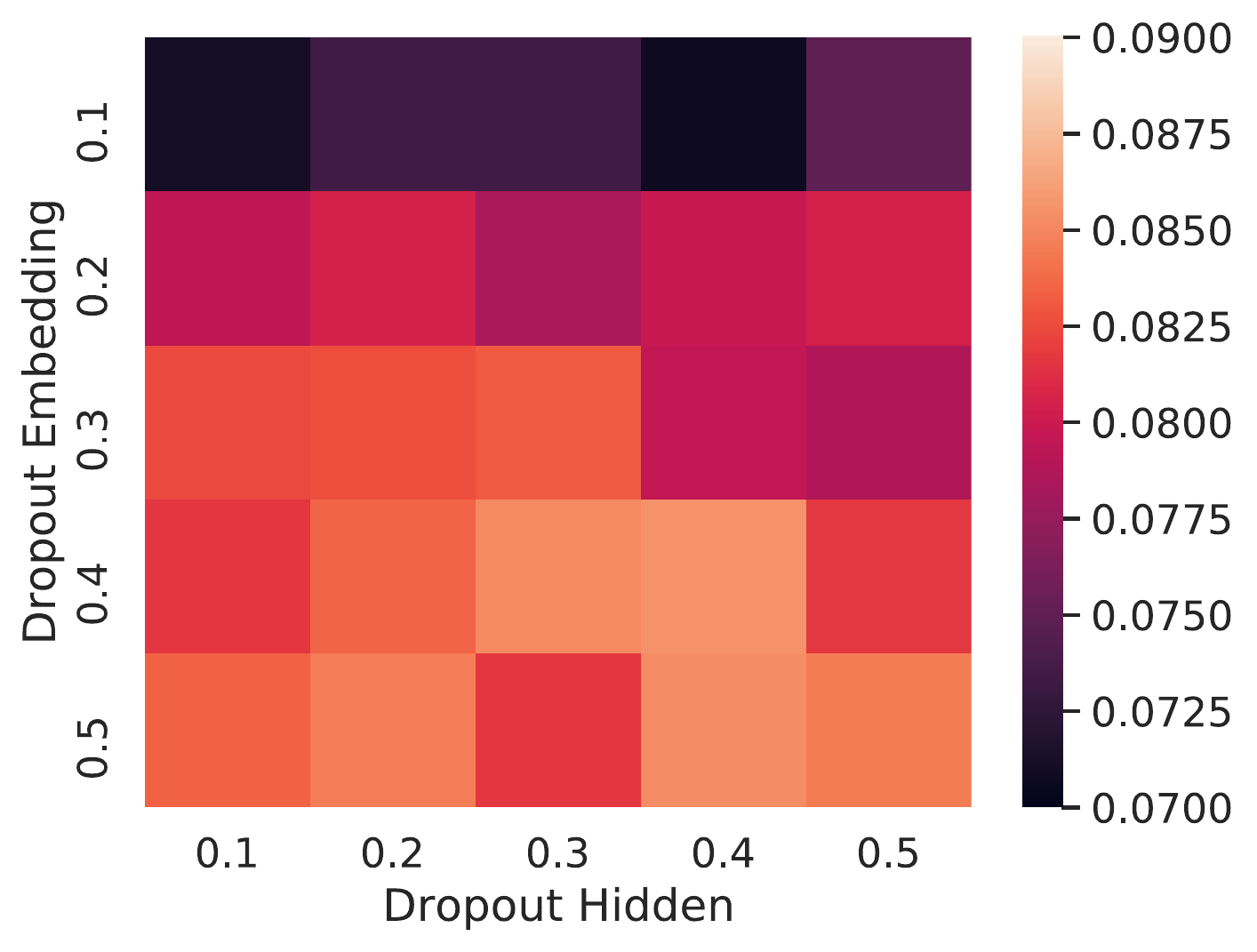}
    }
    \caption{Parameter sensitivity of Dropout ratio on Beauty.}
    \label{fig:beauty-dp}
\end{figure}

\begin{figure}[!h]
    \centering
    \subfigure[HR@5.]{
    \label{fig:cloth-dp-hr5}
    \includegraphics[width=0.465\linewidth]{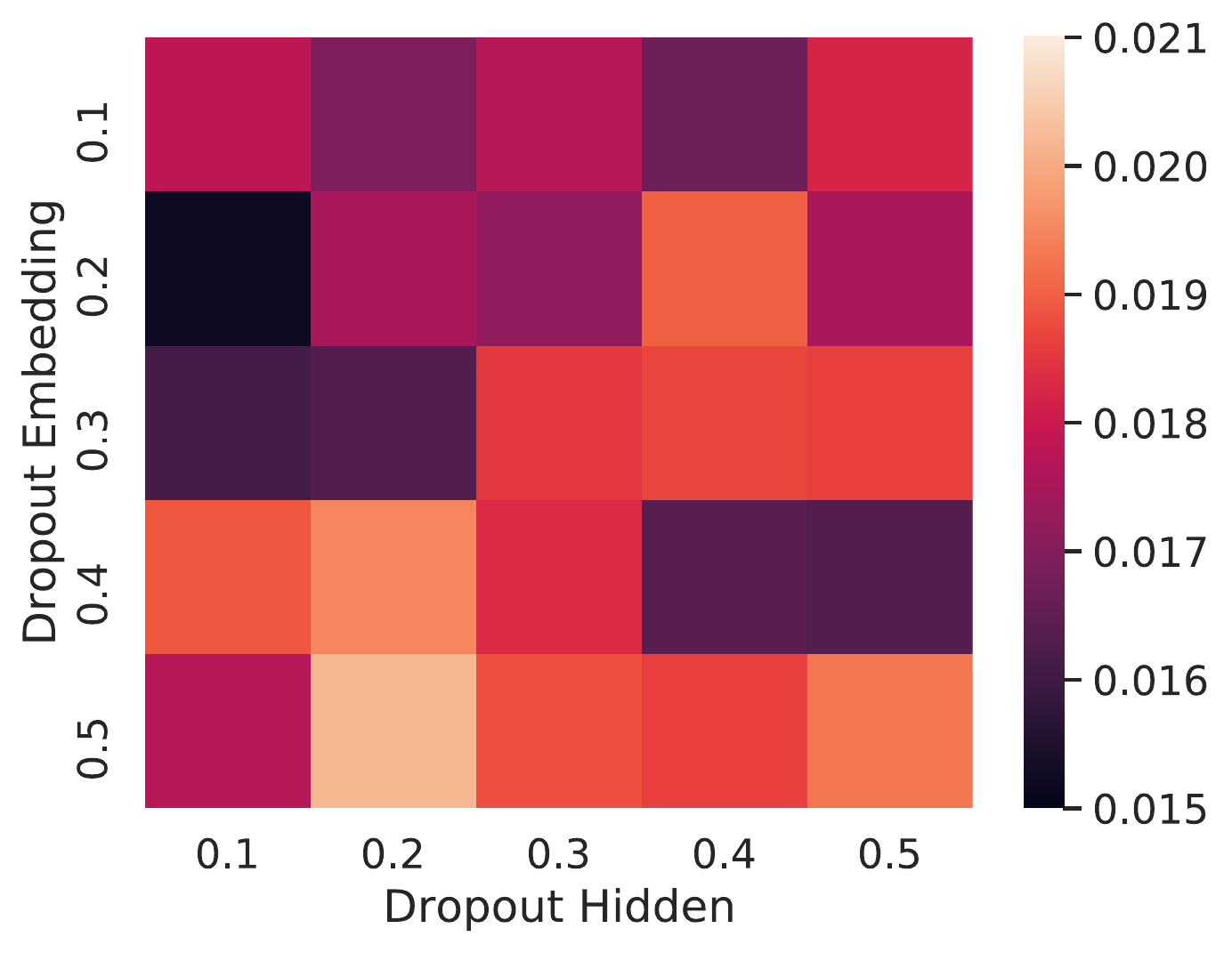}
    }
    \subfigure[HR@10.]{
    \label{fig:cloth-dp-hr10}
    \includegraphics[width=0.465\linewidth]{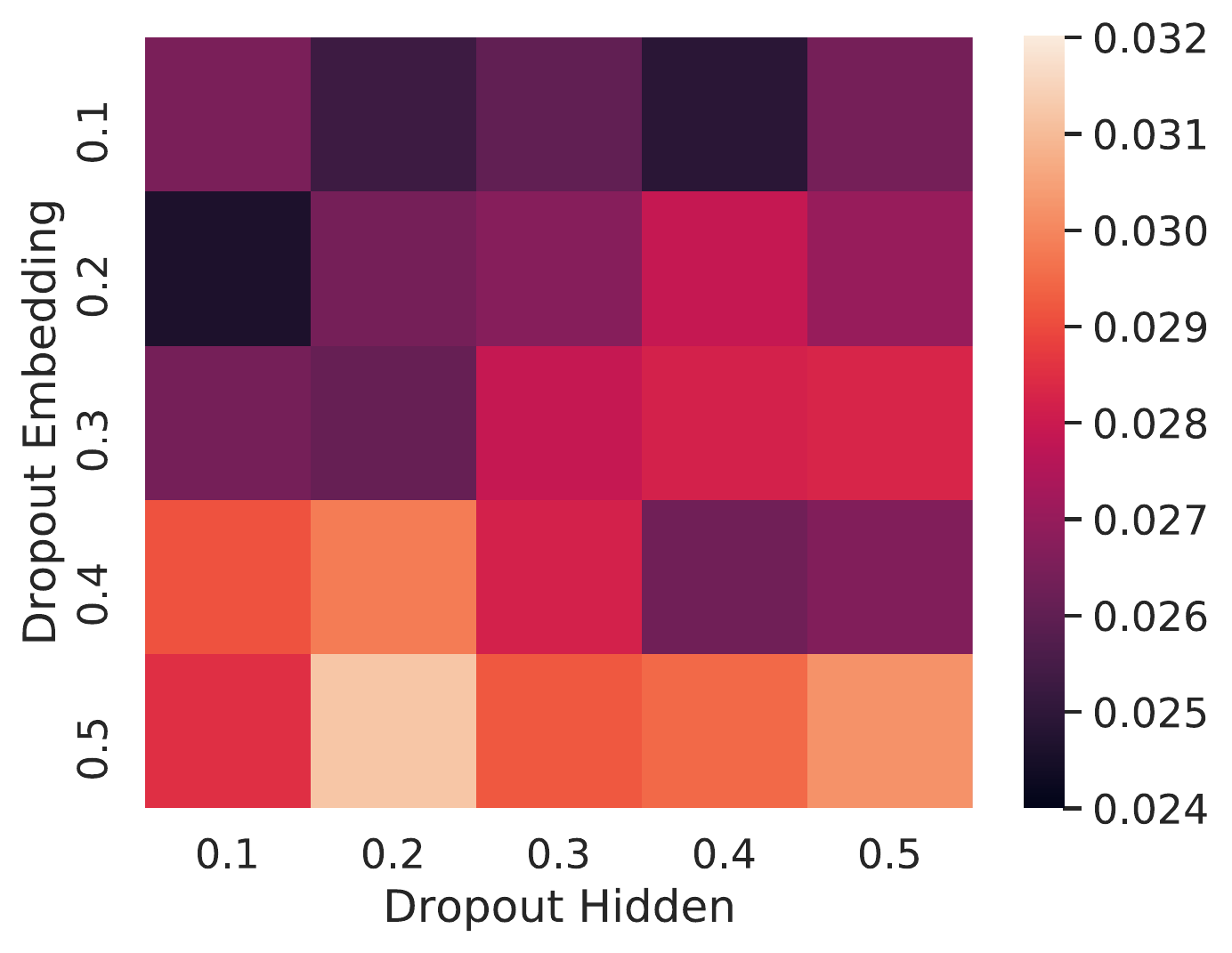}
    }
    \caption{Parameter sensitivity of Dropout ratio on Clothing.}
    \label{fig:cloth-dp}
\end{figure}

\begin{figure}[!h]
    \centering
    \subfigure[HR@5.]{
    \label{fig:sport-dp-hr5}
    \includegraphics[width=0.465\linewidth]{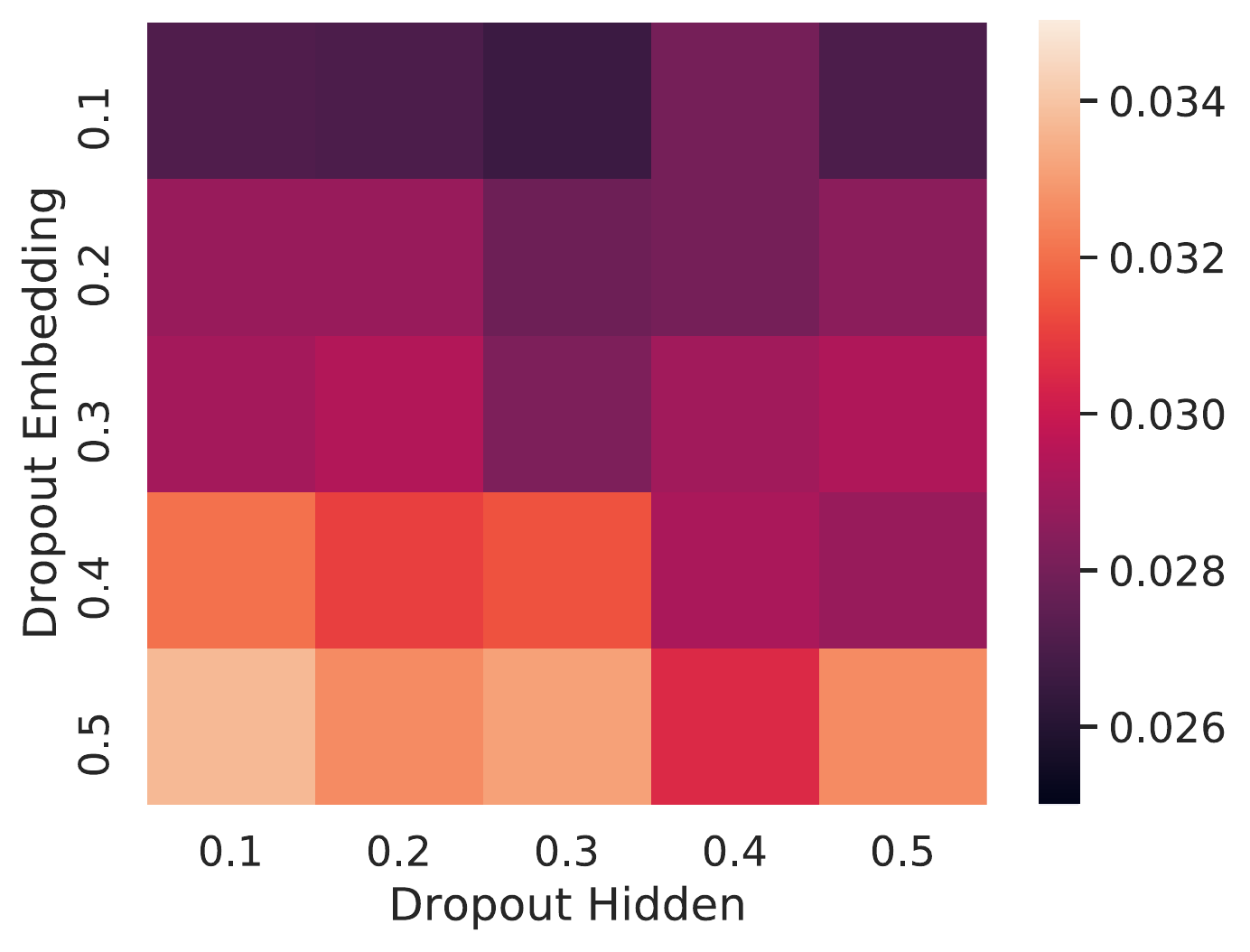}
    }
    \subfigure[HR@10.]{
    \label{fig:sport-dp-hr10}
    \includegraphics[width=0.465\linewidth]{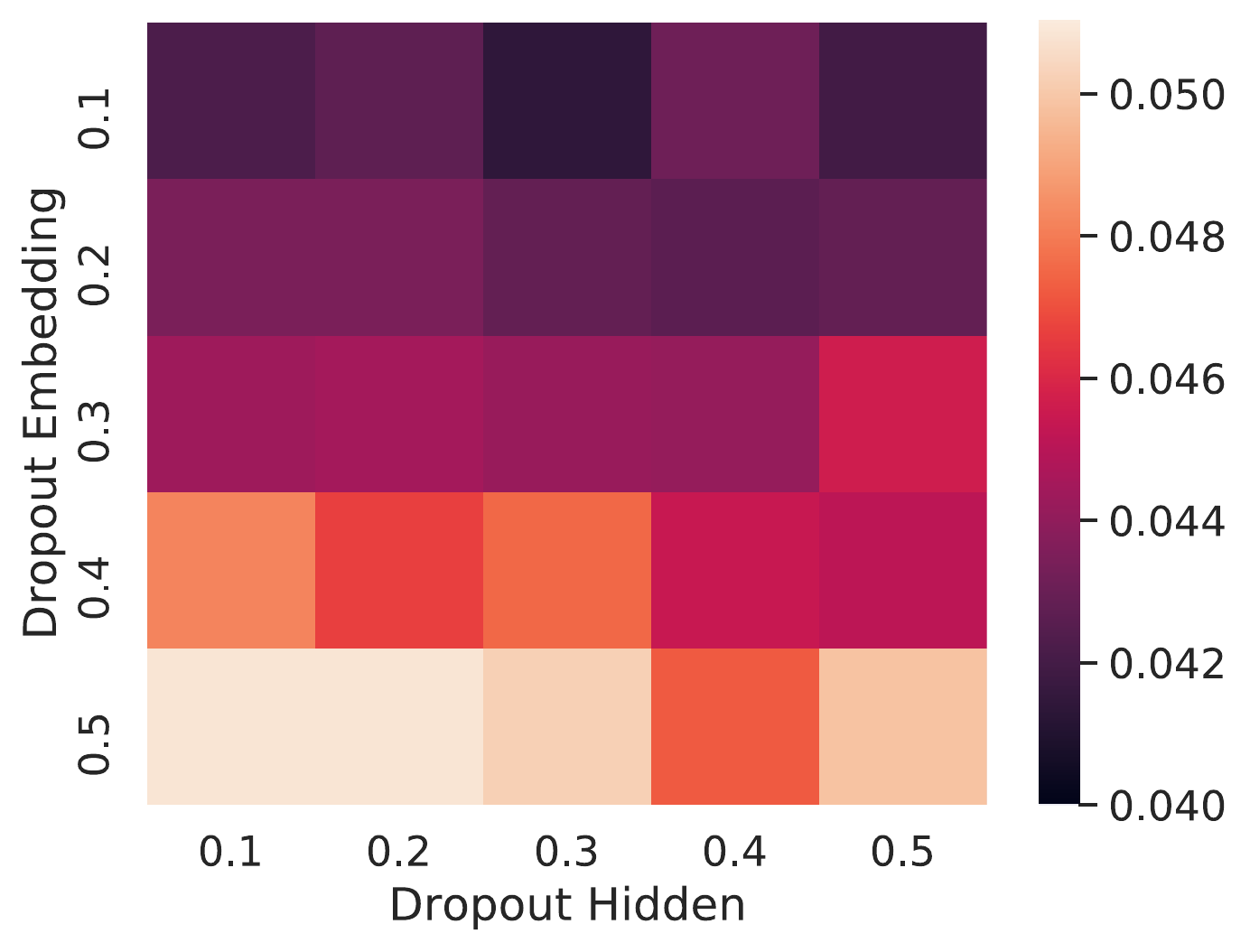}
    }
    \caption{Parameter sensitivity of Dropout ratio on Sports.}
    \label{fig:sport-dp}
\end{figure}

\begin{figure}[!h]
    \centering
    \subfigure[HR@5.]{
    \label{fig:ml-dp-hr5}
    \includegraphics[width=0.465\linewidth]{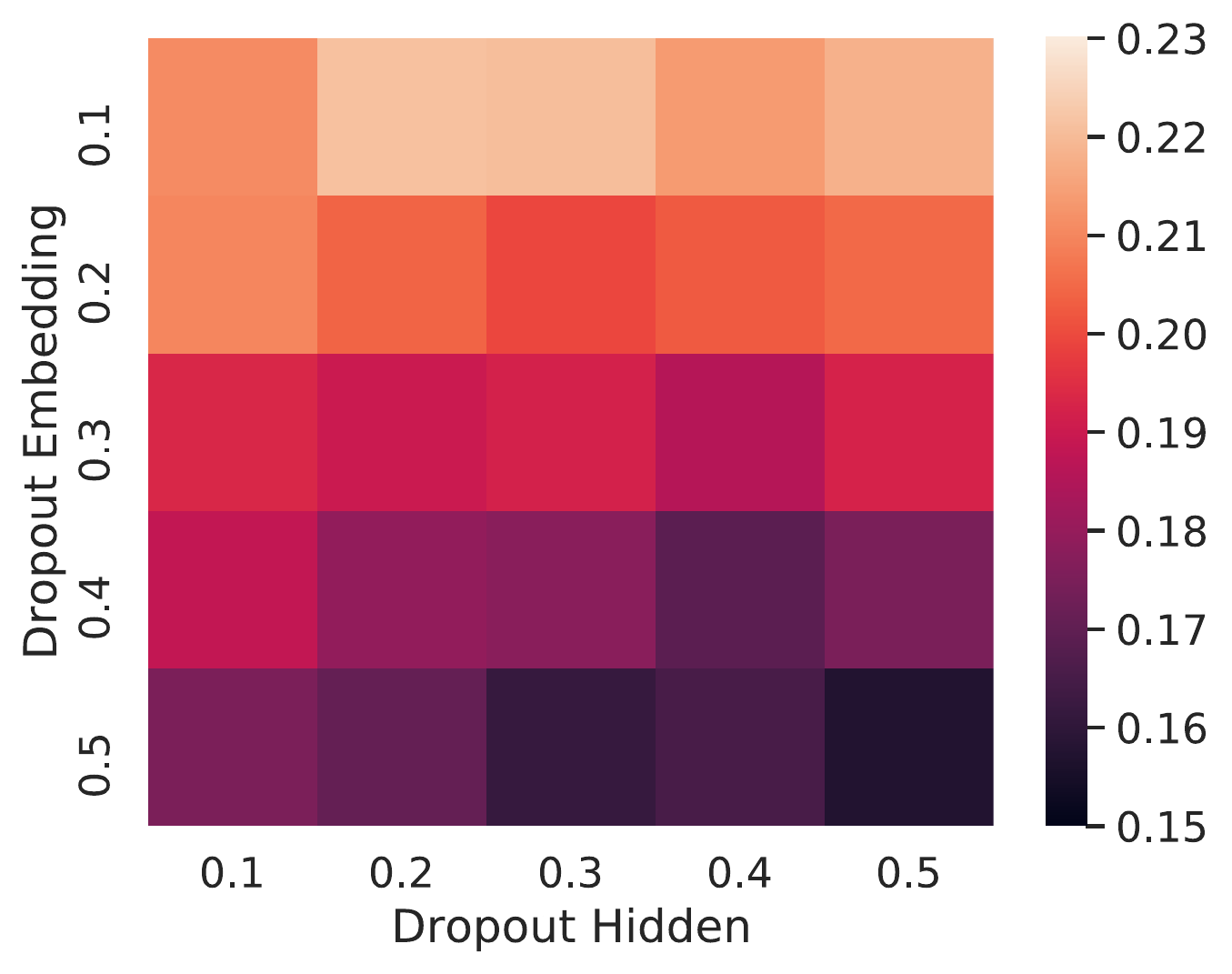}
    }
    \subfigure[HR@10.]{
    \label{fig:ml-dp-hr10}
    \includegraphics[width=0.465\linewidth]{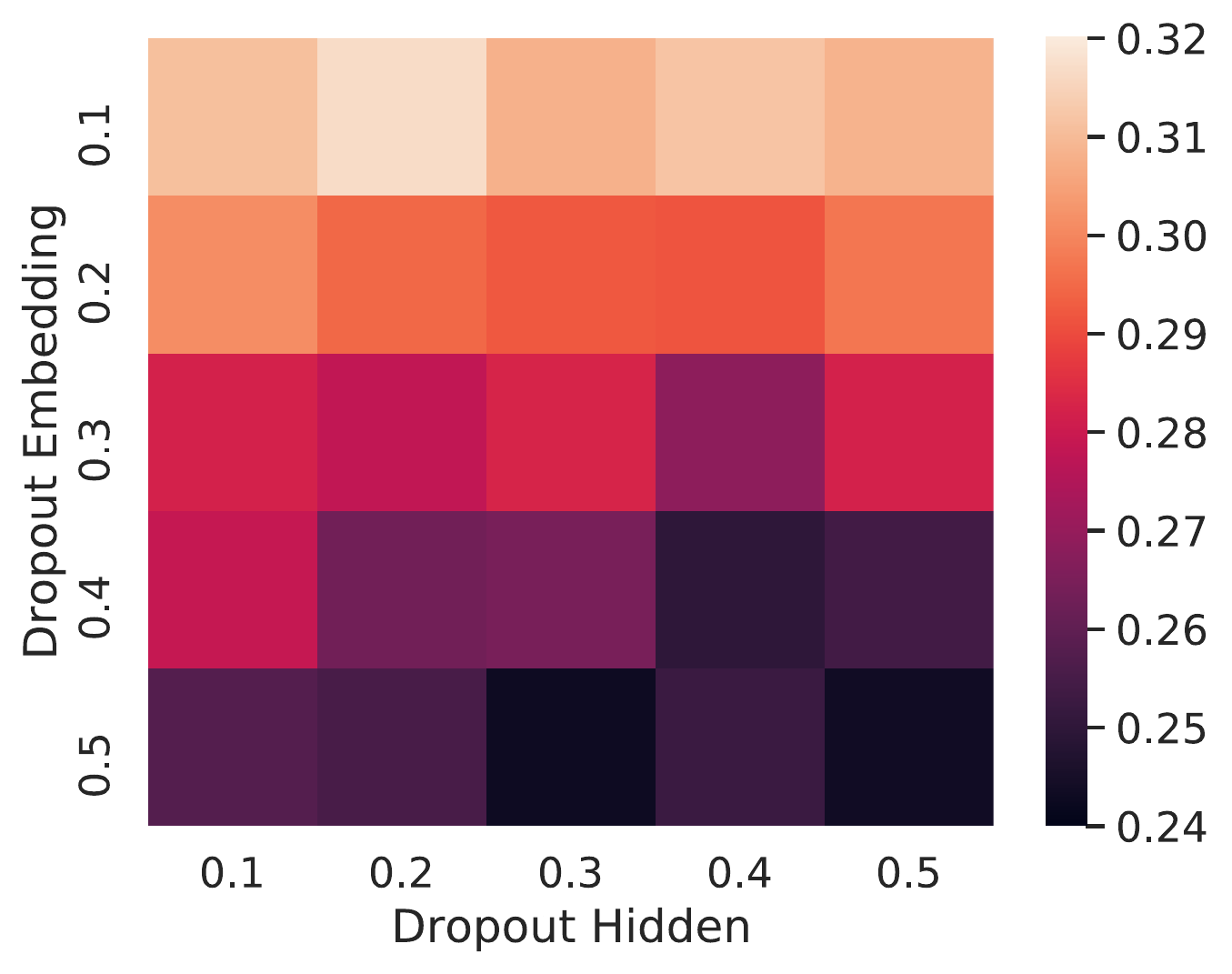}
    }
    \caption{Parameter sensitivity of Dropout ratio on ML-1M.}
    \label{fig:ml-dp}
\end{figure}

\begin{figure}[!h]
    \centering
    \subfigure[HR@5.]{
    \label{fig:yelp-dp-hr5}
    \includegraphics[width=0.465\linewidth]{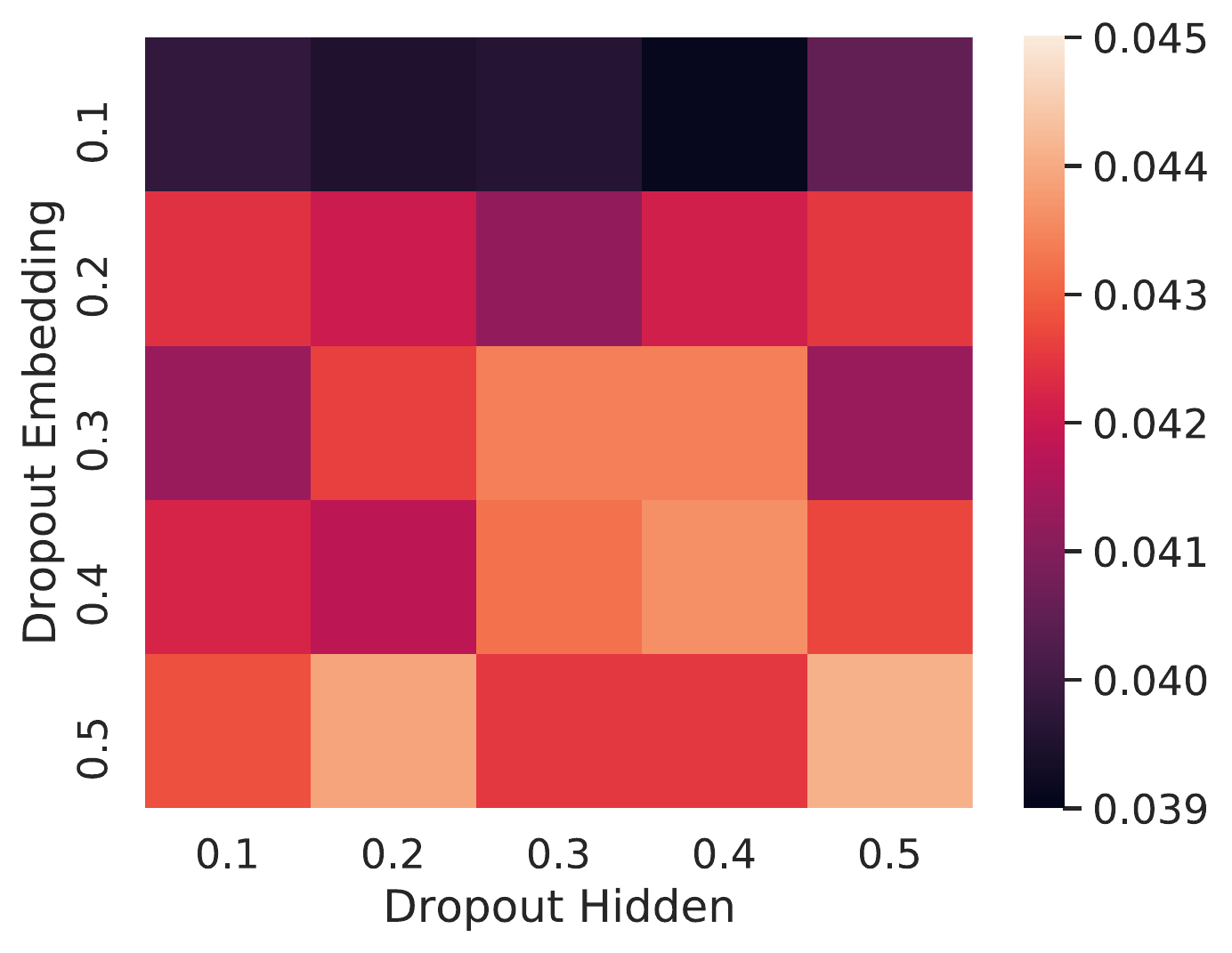}
    }
    \subfigure[HR@10.]{
    \label{fig:yelp-dp-hr10}
    \includegraphics[width=0.465\linewidth]{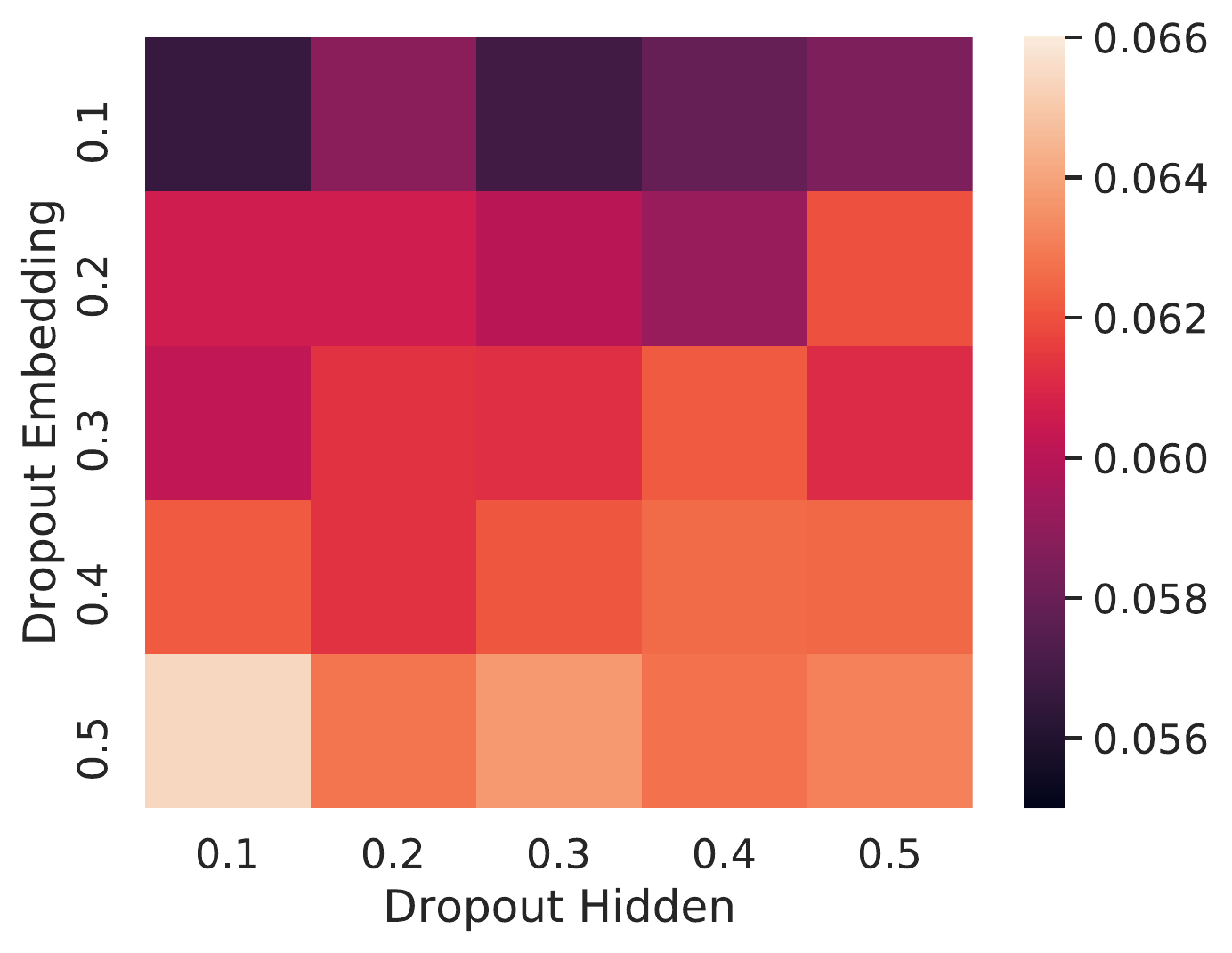}
    }
    \caption{Parameter sensitivity of Dropout ratio on Yelp.}
    \label{fig:yelp-dp}
\end{figure}

Since the Dropout plays an important role in the unsupervised augmentation, the effect of different Dropout strategies are evaluated, which are divided into the Dropout of embedding layer and the Dropout of hidden layers in Transformer. Both Dropout ratios are chosen from $\{0.1,0.2,0.3,0.4,0.5\}$. Results are demonstrated from Figure~\ref{fig:beauty-dp} to~\ref{fig:yelp-dp}.

According to these results that the overall performance is more steady horizontally while fluctuates vertically, the effect of the Dropout of the hidden layers in Transformer has a stronger influence on the overall performance compared with the Dropout of embedding layer. Furthermore, the optimal choice of Dropout strategy on different datasets is different, which is also reflected by Figure~\ref{fig:param-dropout}.

\section{Results of Different Temperatures}
\begin{figure}[!h]
    \centering
    \subfigure{
    \label{fig:tau-hr5}
    \includegraphics[width=0.465\linewidth]{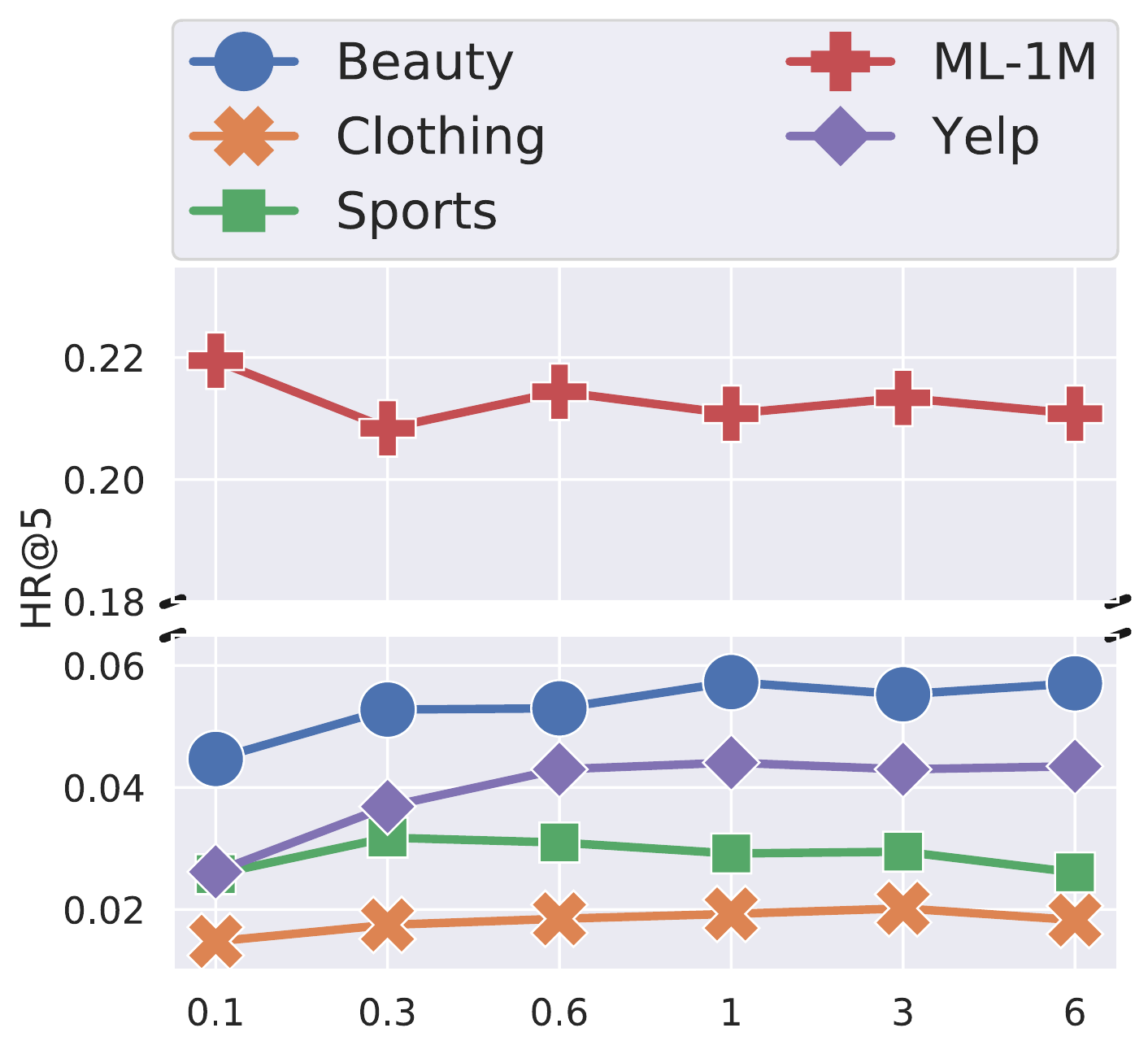}
    }
    \subfigure{
    \label{fig:tau-hr10}
    \includegraphics[width=0.465\linewidth]{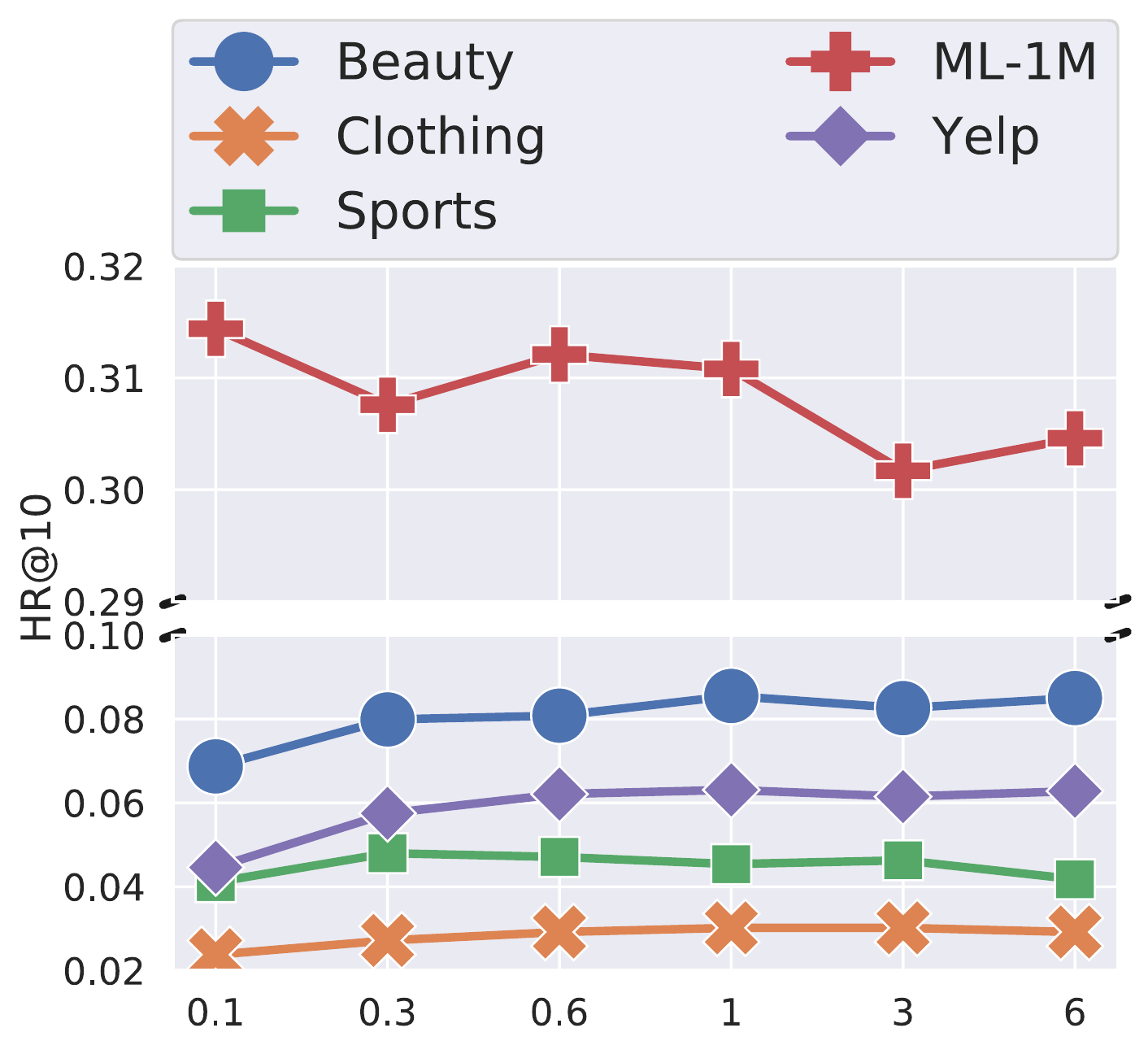}
    }
    \caption{Parameter sensitivity of temperature parameter.}
    \label{fig:tau-param}
\end{figure}

In the calculation of NCE, there is a temperature parameter $\tau$ in Equation (\ref{eq:reg}). This parameter would affect the scale of the estimation of mutual information. The evaluation of the effect of $\tau$ chosen from $\{0.1,0.3,0.6,1,3,6\}$ is presented in Figure~\ref{fig:tau-param}. It can be concluded that within a reasonable range, the overall performance of DuoRec is steady, although there is a slight drop on Beauty and Yelp datasets when $\tau$ is small.

\end{document}